\documentclass[sigconf]{acmart} 
\AtBeginDocument{%
  }

\copyrightyear{2026}
\acmYear{2026}
\setcopyright{cc}
\setcctype{by}
\acmConference[CCS '26] {Proceedings of the 2026 ACM SIGSAC Conference on Computer and Communications Security}{November 15--19, 2026}{The Hague, Netherlands.}
\acmBooktitle{Proceedings of the 2026 ACM SIGSAC Conference on Computer and Communications Security (CCS '26), November 15--19, 2026, The Hague, Netherlands}
\acmISBN{979-8-4007-2871-6/2026/11}
\acmDOI{10.1145/3830454.3846754}

\usepackage{minted}
\usepackage{xcolor}
\usepackage{listings}
\usepackage{enumitem} 
\usepackage{appendix}
\usepackage[ruled,linesnumbered]{algorithm2e}
\usepackage{tabularx}
\usepackage{siunitx}
\usepackage{hyphenat}
\usepackage[table]{xcolor}
\usepackage{cleveref}
\usepackage{pgfmath}
\usepackage{shortcuts}
\usepackage{subcaption}
\usepackage{adjustbox}
\usepackage{multirow}
\definecolor{LightGray}{gray}{0.95}

\newcommand{\PaperTitle}{You Get What You Sample: Evaluating Sampling Strategies for Web Security Measurements}

\begin{document}
\title{\PaperTitle}

\author{Xuenan Zhang}
\email{xuenan.zhang@cispa.de}
\orcid{0000-0002-2124-769X}
\affiliation{%
  \institution{CISPA Helmholtz Center for Information Security}
  \city{Saarbrücken}
  \country{Germany}
}

\author{Yuqing Yang}
\email{yuqing.yang@cispa.de}
\orcid{0000-0002-5149-7913}
\affiliation{%
  \institution{CISPA Helmholtz Center for Information Security}
  \city{Saarbrücken}
  \country{Germany}
}

\author{Giancarlo Pellegrino}
\email{pellegrino@cispa.de}
\orcid{0009-0007-6223-8945}
\affiliation{%
  \institution{CISPA Helmholtz Center for Information Security}
  \city{Saarbrücken}
  \country{Germany}
}

\renewcommand{\shortauthors}{Zhang et al.}

\begin{abstract}

  Web measurement studies rely on domain datasets such as Tranco to quantify the prevalence and impact of security issues at scale, but exhaustively analyzing these datasets is often infeasible because of the cost of advanced analysis techniques, requiring the use of sampling. Despite its widespread use, sampling remains largely guided by convention---most commonly \emph{Top $N$} domain selection---rather than evidence, and its influence on the validity and generalizability of security findings has received little systematic evaluation. Consequently, it remains unclear whether common sampling strategies introduce systematic bias, distort observed vulnerability rates, or limit comparability across studies.

  In this work, we undertake, to the best of our knowledge, the first comprehensive investigation into how sampling methodologies affect the measurements and the conclusions. Through a comprehensive literature review and large-scale measurements of 500k Tranco and 24.8M Common Crawl hosts, we perform a comparative evaluation of datasets and sampling strategies. We show that, while Top $N$ sampling may be a rational strategy, the researchers have to bear in mind that Top $N$ does not reflect the overall distribution of the web. Instead, probability-based strategies yield stable, unbiased estimates for prevalence and many impact objectives. Hybrid sampling provides no advantages over pure probability sampling, as its deterministic prefix consistently contributes negatively to accuracy. Building on these results, we provide data-backed guidance for future studies, proposing to use an adaptive probability-based sampling strategy that remains effective even when the prevalence of the target issue is unknown. \footnote{This is an extended version of the ACM CCS paper (\url{https://doi.org/10.1145/3830454.3846754}).}
    
\end{abstract}




\maketitle

\section{Introduction}
\label{sec:intro}

Web measurement studies play a crucial role in understanding the prevalence and impact of security issues across the Web, relying on seed datasets of millions of domains, such as Tranco~\cite{LePochat2019} and CrUX~\cite{chrome_crux}, or public crawled web archives such as Common Crawl~\cite{commoncrawl}. Analyzing these datasets in full is often infeasible due to high computational cost, as measurements may require crawling and applying costly program analysis techniques such as dynamic taint tracking~\cite{foxhound}, SMT solvers~\cite{steffens2020pmforce}, or static analysis~\cite{khodayari2021jaw,khodayari2023s,khodayari2024great}. As a result, web measurements commonly rely on sampling, analyzing only subsets of domains to reduce evaluation cost.

However, the extent to which measurement findings generalize to the broader Web depends on two closely related design choices: \emph{which population of domains is used as the measurement basis}, and \emph{how domains are sampled from that population}. The first choice concerns the seed dataset. Popularity-ranked lists such as Tranco~\cite{LePochat2019} emphasize highly visited domains and are therefore a natural fit for impact-oriented studies, whereas broad web archives such as Common Crawl~\cite{commoncrawl} aim to cover the Web more comprehensively and are better tailored with prevalence-oriented analyses. The second choice concerns the sampling strategy applied to that dataset. In current web measurement practice, the dominant approach is \emph{Top $N$} sampling, in which only the top 1k--10k domains are analyzed, for example in studies of domain takeover~\cite{squarcina2021can}, deployment of web security mechanisms~\cite{poteat2021you,rautenstrauch2023leaky,khodayari2023s}, or fingerprinting attacks~\cite{zhao2024towards,topcuoglu2024untangle}. Other strategies, such as random, stratified, and hybrid sampling~\cite{al2023large,munir2023cookiegraph}, appear far less frequently. Consequently, key measurement decisions are often made by convention rather than by evidence, even though mismatches between dataset, sampling strategy, and research objective can introduce systematic bias and distort conclusions about security issues on the Web.

As such, evaluating methodological choices in empirical research is crucial to distinguish real security phenomena from artifacts introduced by measurement design. Prior work has focused on multiple parts of the web measurement pipeline, including ranking construction~\cite{LePochat2019,xie2022building,ruth2022toppling}, browser configurations when crawling~\cite{demir2023similarity}, crawling strategies~\cite{stafeev2024sok}, and tool behavior~\cite{ahmad2020apophanies}, yet sampling remains comparatively underexplored, despite determining which parts of the Web are observed at all. To date, we lack evidence on whether common sampling methodologies systematically over- or underrepresent different parts of the Web. If they do, reported vulnerability rates may be skewed, security mechanisms may appear more---or less---deployed than they truly are, and findings across studies may not be meaningfully comparable. More fundamentally, we still lack a systematic understanding of which sampling strategies are reliable, when they fail, and how dataset choice interacts with sampling to affect security measurements. This leaves open a fundamental question: \emph{how much do sampling methodologies affect the measurements we report and the conclusions we ultimately draw?}

In this paper, we address this question through systematic investigations on sampling methodologies in web security measurements. First, we map the landscape of sampling methodologies used in web security measurement by conducting a comprehensive review of prior work and by incorporating established sampling techniques from statistics, resulting in eight distinct strategies. Second, we perform a large-scale, ground-truth evaluation of these strategies covering multiple classes of security issues including client-side XSS, HTTP security header misconfigurations, and TLS configuration errors, using a dataset of 500k domains drawn from the Tranco ranking~\cite{LePochat2019} and 24.8M hosts from the Common Crawl archives~\cite{commoncrawl}. Third, we analyze how and why sampling strategies behave as they do, decomposing hybrid approaches into their constituent components, and evaluating sampling effectiveness along multiple dimensions—prevalence, impact, error magnitude, stability, and sensitivity to the underlying distribution of vulnerabilities.

Our findings reveal clear and consistent patterns. Top $N$ sampling, despite being the dominant strategy in prior work, deviates from the overall distribution of its target population: in our impact measurements, it can over- or under-estimate the full-population value by up to 10.6 percentage points in datasets with smaller sizes of 10k domains, and simply sampling more domains does not remove this bias.
Probability-based methods behave very differently: on the 500k-domain Tranco dataset, Random, Systematic, and Stratified sampling stay within a narrow margin of the true value, making their estimates effectively stable for the measurement goals. Hybrid strategies do not solve the Top $N$ sampling problem: they inherit its bias and spend much of their remaining sample budget correcting it. We further show that sampling correctly from the wrong population still yields the wrong answer. 
{For web-wide prevalence over a broader host population, Tranco may not be the most suitable dataset: Random sampling on Tranco remains up to 19.6 percentage points away from the Common Crawl estimate, with a systematic average underestimation rate of 30\% even when sampling size increases.}
%
%
Finally, we propose a new practical sampling strategy, the \emph{Adaptive Probability Sampling}: begin with a 1--2\% probability-sampling pilot and expand adaptively, as most non-sparse issues stabilize by 5--10\% sampling. Overall, our study highlights the limitations of the most popular Top $N$ sampling strategy, and provides practical guidance for researchers to select sampling strategies.

\noindent\textbf{Contributions.} This paper makes the following contributions:

\begin{itemize}[leftmargin=1.2em]
    \item We present, to the best of our knowledge, the first empirical study to systematically evaluate how sampling strategies affect security measurements on the Web.
    \item We identify eight sampling strategies (six used in prior web security research and two drawn from statistical sampling theory) and evaluate both their basic and hybrid variants.
    \item We conduct a large-scale ground-truth measurement of 500k domains from Tranco and 24M from Common Crawl to quantify sampling error across multiple security issues, measurement objectives (prevalence and impact), and evaluation metrics.
    \item We provide a decomposition of hybrid sampling using Shapley analysis, isolating the contributions of deterministic and probabilistic components and revealing structural biases in widely used designs.
    \item We deliver data-backed, practical guidance for future researchers, including recommendations on which sampling strategies to use, how large samples should be, and how to make these decisions when the prevalence of the target issue is unknown.
    \item We distill six lessons learned about sampling behavior that generalize across strategies, datasets, and measurement goals.
\end{itemize}

\section{Background}
\label{sec:background}

Before presenting our research questions, we provide essential background, beginning with an overview of sampling strategies, followed by representative problems studied in web security measurement, and concluding with common measurement objectives.

\subsection{Sampling Strategies}

We can broadly distinguish three families of methods~\cite{berndt2020sampling, etikan2017sampling}:

\subsubsection*{1) Probability Sampling.} A set of sampling methods in which every unit in the population (e.g., a domain in a ranking such as Tranco) has a \emph{known, non-zero probability of selection} prior to sampling, and units are selected through a randomized process respecting those probabilities. Common methods include \emph{Simple Random Sampling}, equivalent to the Random N strategy used in web measurement research. Other examples include \emph{Systematic Sampling}, where samples are selected at fixed intervals (e.g., every i-th element) after a random starting point, and \emph{Stratified Random Sampling}, where the population is first partitioned into subgroups (strata) and samples are randomly drawn from each stratum proportional to its size.

\subsubsection*{2) Non-Probability Sampling.} A family of methods where a sample is not drawn through a fully randomized mechanism that guarantees every unit a measurable chance of selection. One example in web measurement is \emph{Top $N$} selection, where only the highest-ranked $N$ domains are chosen deterministically, leaving all other domains with zero probability of being sampled.

\subsubsection*{3) Hybrid Sampling.} Hybrid Sampling combines elements of both probability and non-probability methods, e.g., selecting a deterministic Top N prefix and then applying a randomized sampling method to the remaining population.

\subsection{Commonly-studied Web Security Issues}

A wide range of security issues have been measured on the web, spanning web vulnerabilities, deployment of defenses, and insecure configurations. In this section, we present three representative categories that are frequently analyzed in the wild: (i) \emph{client-side vulnerability detection}, exemplified by client-side XSS; (ii) \emph{HTTP header configurations}, where improper or missing headers weaken browser defenses; and (iii) \emph{TLS configuration issues}, where protocol or certificate misconfigurations undermine secure communication.

\subsubsection*{Client-Side Vulnerability Detection.} Modern websites depend on complex client-side logic and third-party scripts, making browser-side vulnerabilities an interesting target for large-scale web measurements, such as \emph{client-side XSS}~\cite{lekies201325, melicher2018riding, steffens2019don, rautenstrauch2024auth}, \emph{DOM clobbering}~\cite{khodayari2023s}, and \emph{client-side request hijacking}~\cite{khodayari2024great}. 
\emph{Client-side XSS} is one of the most studied classes, and it occurs when untrusted data reaches JavaScript execution sinks (e.g., DOM APIs) without sanitization. Unlike server-side XSS, which requires injecting payloads into server contexts and raises ethical risks, client-side XSS can be detected purely via browser-side instrumentation, enabling safer, scalable measurements.

\subsubsection*{HTTP Security Header Configurations.} Another interesting subject for web measurements are HTTP security headers. Headers are convenient as they let websites enforce browser-side protections without modifying client code, mitigating a wide range of threats such as XSS, cookie theft, and protocol downgrades.

Prior measurements focused on script and resource controls (\texttt{Content-Security-Policy}~\cite{calzavara2016content}), framing defenses (\texttt{X-Frame-Opt\allowbreak ions})~\cite{huang2012clickjacking}, MIME type security (\texttt{X-Content-Type-Options})~\cite{barua2011server}, secure cookies (\texttt{Secure}, \texttt{HttpOnly}, and \texttt{SameSite})~\cite{kwon2019security}, and HTTPS enforcement (\texttt{Strict-Transport-Security})~\cite{dolnak2017introduction}.


\subsubsection*{TLS Misconfigurations.} TLS is another core pillar of web security and is a major focus of web measurements. Prior work analyzes common deployment flaws, including invalid or expired certificates~\cite{chung2016measuring}, domain mismatches~\cite{akhawe2013here}, weak or deprecated protocol versions (e.g., TLS 1.0/1.1)~\cite{manfredi2019lost}, insecure cipher suites~\cite{calzavara2019postcards}, and improper key exchange or handshake configurations~\cite{bhargavan2013implementing}. 

\subsection{Objectives of Measurement Studies}

Web security measurements are a key method for studying security issues at Internet scale rather than through isolated case studies. Common goals include quantifying prevalence, identifying emerging threats, validating assumptions, comparing defenses, and tracking ecosystem evolution. In this paper, we discuss two common measurement objectives, i.e., prevalence and impact, while noting that empirical studies often serve both simultaneously. Rather than proposing a taxonomy, we use these objectives as a conceptual framework to analyze how sampling decisions affect the validity, implications, and generality of web measurements.

\subsubsection*{Prevalence Measurement.} Prevalence studies estimate how widespread a security property or vulnerability is across a typically large population of websites. This objective answers questions about adoption, exposure, or susceptibility rates, drawing conclusions about the broader web from a sampled set of domains. For example, Khodayari et al.~\cite{khodayari2023s} measure the prevalence of DOM clobbering vulnerabilities and the effectiveness of deployed defenses across the Tranco Top 5K.

\subsubsection*{Impact Measurement.} Impact studies quantify the real-world consequences of a security issue beyond its mere presence. This includes estimating affected users, data exposure, behavioral influence, or ecosystem-level effects from the popularity of the domains considered. Unlike prevalence, which measures \emph{how many sites are affected}, impact measurements emphasize \emph{how much the issue matters}, often prioritizing smaller sets of high-profile domains (e.g., top-ranked sites) where security failures have disproportionate reach and user impact.

\section{Problem Statement}

The overarching goal of this paper is to understand how sampling choices shape the results of web security measurements and, ultimately, the scientific conclusions drawn from them. Although sampling is a fundamental step in large-scale measurement, the assumptions behind existing strategies, and their impact on accuracy and bias, are often left implicit. Our work aims to make these assumptions explicit, quantify their effects, and provide practical guidance for researchers who must select sampling methods under real resource constraints. To structure our analysis, we articulate the following research questions:

\subsubsection*{\textup{\textbf{RQ1: What sampling strategies are used in practice, and what alternatives should be considered?}}}
We begin by identifying the sampling strategies that appear in the web measurement literature, characterizing both their conceptual design and their practical motivations. This includes deterministic strategies such as Top~N, probability-based methods such as Random, Systematic, and Stratified sampling, and hybrid designs combining ranking-based and probabilistic components. By surveying existing practice and formalizing the space of possible approaches, we establish the landscape of strategies that must be evaluated (\Cref{sec:sample}).

\subsubsection*{\textup{\textbf{RQ2: How do different sampling strategies behave when applied to real-world web security measurements?}}}
The core contribution of our study is an empirical, large-scale comparison of sampling strategies. We examine not only their accuracy but also their stability, convergence behavior, and the structural assumptions embedded in their design. This includes understanding when strategies fail, whether their errors shrink with more data, and how hybrid methods decompose into the contributions of their deterministic and probabilistic components. These analyses allow us to quantify how sampling choices influence measurement outcomes and reveal previously unexamined limitations in widely used approaches (\Cref{sec:sampleeval}).

\subsubsection*{\textup{\textbf{RQ3: How do different datasets impact sampling results when measuring security impacts or prevalence?}}}
In this study, we further examine how sampling from different datasets may affect the results of security measurements. In security measurement, the goal can be categorized into two types: measuring how many top-visited domains are affected by a security issue (impact), and measuring how frequently a security issue occurs on the Web at large (prevalence). In particular, we evaluate the deviation of sampling results by applying the same sampling strategy on different datasets, namely Tranco (ranked based on popularity) and Common Crawl (not popularity-ranked like Tranco). We analyze the overall distribution of both datasets and evaluate how random sampling from these datasets may lead to over- or under-estimation, and what is the level of expected deviation. This analysis allows us to understand how the choice of dataset may impact the results of security measurements, and how sampling strategies perform on different datasets (\Cref{sec:prevalence}).

\subsubsection*{\textup{\textbf{RQ4: How should researchers choose sampling strategies and sample sizes when the prevalence of the issue is unknown?}}}
In practice, researchers rarely know how common a security issue is when designing a study. They must choose a sampling method and a sampling budget under uncertainty, while avoiding both overshooting (expending excessive resources on a problem that would require less) and undershooting (collecting too little data to observe rare events). Our final research question focuses on deriving practical guidance from our results: identifying which sampling functions are robust in the absence of prior knowledge, how sample size requirements scale with prevalence, and how to design sampling plans that remain cost-effective across a wide range of possible scenarios (\Cref{sec:bestpractice}).


    

\section{Literature Review for Sampling Strategies}
\label{sec:sample}

\begin{table*}[ht!]
\centering
\small
\begin{tabular}{ l p{11.2cm} l c} 
\toprule
\textbf{Name} & \textbf{Description} & \textbf{Type} & \textbf{No. Papers} \\ 
\midrule

\textbf{Top N} &
Top N sampling obtains a ranking list based on metrics such as popularity and samples the first $N$ websites to form the dataset. Among the 107 sampling papers, 87 use this strategy, making it the most common. &
Basic &
87\\ 

\textbf{Random N} &
Randomly samples multiple websites from the total dataset, regardless of popularity. Used by 7 papers. &
Basic &
7\\ 

\textbf{Top N plus Random} &
First selects the $N$ most popular websites, then randomly selects $M$ websites from the remaining population, forming a dataset of $N+M$ websites. Used by 4 papers. &
Hybrid &
4\\ 

\textbf{Top N plus Stratified} &
First selects the $N$ most popular websites, divides the remaining websites into $t$ strata of varying sizes (e.g., ranks 100--1{,}000, 1{,}000--10{,}000), and randomly selects $M$ websites from each stratum, forming a dataset of $N+tM$ websites. Used by 7 papers. &
Hybrid &
7\\ 

\textbf{Top N plus Bucket} &
First selects the $N$ most popular websites, divides the remaining websites into $t$ equal-sized buckets, and randomly selects $M$ websites from each bucket, forming a dataset of $N+tM$ websites. It differs from stratified sampling only in that the buckets are equal-sized. Used by 1 paper. &
Hybrid &
1\\ 

\textbf{Random K from Top N} &
Randomly selects $K$ websites from the top $N$ websites, forming a dataset of size $K$. Used by 1 paper. &
Hybrid &
1\\
\midrule

\textbf{Systematic sampling} &
Selects samples at fixed intervals (incl. fractions), e.g., every 5th website or from an ordered list, usually starting from a random point.&
Basic &
\cite{etikan2017sampling} \\ 

\textbf{Stratified random sampling} &
Divides the population into strata (subgroups), often based on website rankings, where lower-ranked websites form larger strata. Randomly samples within each stratum. &
Basic &
\cite{etikan2017sampling}\\ 

\bottomrule
\end{tabular}
\caption{Sampling strategies from our literature review.}
\label{tab:sampling_from_lit}
\end{table*}

\subsubsection*{Literature Review.} 

To answer RQ1, we collected all papers published in six top security and measurement venues, i.e., IMC, CCS, USENIX Security, NDSS, IEEE S\&P, and WWW. We selected these papers between 2020 and 2024 for two reasons. First, we consider the shifting landscape of domain popularity lists, as Tranco (since late 2019) and CrUX have become common alternatives after Alexa~\cite{alexa_top1m} retired in May 2022. Second, web domains emerge and may become outdated. Therefore, capturing recent works reduces the risk of domains being taken down or inaccessible.

The analysis consists of two phases. First, we filter out the papers unrelated to web measurement by reading the titles and abstracts, resulting in 185 papers. Then, we analyze these papers in detail, resulting in 107 papers. These papers sample from a variety of datasets, with Tranco (61) and Alexa (41) being the most dominant datasets, followed by CrUX (9), Common Crawl (7), Cisco Umbrella (3)~\cite{cisco_umbrella}, and SecRank (2)~\cite{xie2022building}. For Quantcast~\cite{quantcast_topm}, Curlie~\cite{curlie}, Open PageRank~\cite{openpagerank}, Majestic~\cite{majestic}, and Cloudflare Radar~\cite{cloudflare_radar}, only one paper is associated. In addition, three hostname-enumeration datasets (CAIDA DNS Names~\cite{caida_dns_names}, Censys CT~\cite{censys_ct}, and Rapid7 FDNS~\cite{rapid7_fdns}) each appear in one paper. Based on this analysis, Tranco is the most frequently used popularity-ranked domain list, while Common Crawl is the most frequently used broad crawl-derived dataset that is not ordered by popularity in the same manner as Tranco. As such, we use these two datasets as the source of our sampling experiments. On top of that, we utilize the recent Tranco list to improve representativeness, as CrUX and Cloudflare Radar rankings have been integrated into the default Tranco list since August 1, 2023~\cite{tranco_home}.

In addition to the six strategies identified via the literature review on web security measurements, we also included sampling strategies commonly used in statistics and applied mathematics to identify possible strategies that have been overlooked by the existing web measurement literature but still bear potential to perform representative sampling that can be advocated. Our source for this part of the survey is Etikan et al.~\cite{etikan2017sampling}\footnote{This source is the most cited paper when searching ``sampling methods'' and ``sampling strategy'' on Google Scholar.}. We thus identified two additional sampling strategies. The results of our survey are in \Cref{tab:sampling_from_lit}. There are two basic types: Top N and random sampling, and 4 hybrid strategies used by related papers by combining two strategies in different ways. We observe that the Top N is mostly used in related literature, whereas hybrid strategies generally are less popular. For comprehensiveness, we will evaluate all these 8 strategies in the rest of the paper.

\subsubsection*{Sampling Parameters.} For strategies requiring explicit parameters, we used configurations drawn from prior work whenever possible. While most methods only need a target sample size, stratified, bucket, and systematic sampling derive their sample size from other parameters. For stratified and bucket sampling, we adopted the setups in~\cite{strata} and~\cite{hantke2023you} respectively. As our survey found no prior use of systematic sampling, we set its interval heuristically as the ratio between the desired sample size and the total population size.

\section{Datasets}

\subsection{Data Collection}

We collected data from two sources, the Tranco list~\cite{trancolist} and Common Crawl~\cite{commoncrawl}, using a 13-server cluster with 128 physical cores and 2\,TB RAM per machine. We chose these sources to support two distinct measurement goals. Tranco provides a popularity-ranked domain list, which is a common basis for impact-oriented analyses. In contrast, Common Crawl computes a search-engine style ranking based on links between web hosts and uses this ranking to allocate the crawl budget for each host~\cite{commoncrawlwebgraph,commoncrawlhostindex}. Although Common Crawl cannot be assumed to provide a complete snapshot of the entire Web, its substantially larger scale and its ranking mechanism, which is not directly based on user popularity, may allow it to reach further into the long tail of less-popular websites than Tranco, making it better suited for our prevalence-oriented analyses. As a website may contain multiple pages, to determine the best configuration to cover as many pages as possible while maintaining reasonable processing performance, we performed a test experiment on a small dataset comprising 10,000 domains. Our experiment shows that depth plays an important role in coverage. When increasing the depth from 1 to 2, the coverage significantly increases, but when the depth increases to 3, the benefit is marginal. As such, we crawl each domain with a Foxhound-based crawler~\cite{foxhound} with depth of 2 and a maximum of 250 pages per domain, with a timeout of 30 seconds.

\paragraph{Tranco List.}
For impact-oriented measurements, we used the Tranco list~\cite{trancolist} (version 8LZ3V, 24 September 2025) and selected the top 500k domains. We resolved them via Google Public DNS~\cite{google_public_dns} and Cloudflare DNS~\cite{cloudflare_dns}, using batched queries to avoid stressing resolvers. This yielded 435{,}868 resolvable domains. We then partitioned the dataset into 50 shuffled buckets of 10{,}000 domains for parallel crawling across our cluster. The crawl ran from 24 September to 8 October 2025. For each visited page, we recorded HTTP headers, tainted flows relevant to client-side XSS, and TLS issues. We retained only same-origin, document-type responses with status code 200, and marked a domain as vulnerable if any such page exhibited a misconfiguration. Additionally, if there is a \textit{page} with security issues, then we label the exact \textit{domain} as having corresponding security issues. Overall, 194{,}794 domains showed at least one security issue. 

We also logged collection failures. The most common were DNS \texttt{NXDOMAIN} and server unresponsiveness. Across 10k buckets, the share of successfully acquired domains ranged from 72\% to 96\%. \Cref{app:errortypes} presents our error analysis.

\paragraph{Common Crawl.}
For prevalence-oriented measurements, we used Common Crawl~\cite{commoncrawl}. We retrieved HTTP responses from the Web ARChive (WARC) files of the October 2025 snapshot (ID: \texttt{CC-MAIN-\allowbreak 2025-43})~\cite{commoncrawl_october_2025}. Such a snapshot contains 2.6 billion web pages from 47 million hosts, comprising 468 TB of uncompressed content. As such, we prioritize the feasibility of the processing by distributing WARC files to 50 parallel workers. Also, as each WARC file comprises varying amount of domains, a worker is configured to process at most 25 million URLs. In the end, we processed a total of 1.2B unique URLs and 24{,}834{,}442 hosts, covering approximately 53\% of the reported hosts.


From these records, we extracted HTTP response headers for large-scale analysis. As an archival corpus, Common Crawl supports only analyses over static artifacts. We therefore use it for header-based prevalence measurements and reserve runtime analyses, such as taint-tracked XSS detection and TLS error collection, for the Tranco crawl. Overall, 13{,}594{,}429 of the 24{,}834{,}442 (i.e. 54.7\%) hosts contained at least one misconfigured header.

\paragraph{Ethics.}
We followed standard practices to minimize real-world impact; details are provided in \Cref{sec:ethics}.

\subsection{Security Analyses}
\label{sec:sec_analysis}

We next describe the security analyses applied to the collected data.

\subsubsection{Client-Side XSS Analysis.}
Client-side XSS occurs when tainted JavaScript data flows from a client-side source (e.g., an API reading the navigation URL) to an execution sink (e.g., \texttt{eval}). We consider both \emph{reflected} and \emph{stored} variants and detect them via taint analysis. We collect dynamic data flows with Foxhound~\cite{foxhound}, a Firefox-based browser with dynamic taint tracking, driven through Playwright~\cite{playwright}. Because tainted flows alone do not establish exploitability, we generate and validate attack payloads using the exploit generator of Steffens et al.~\cite{steffens2019don}, following prior methodology~\cite{steffens2019don, lekies201325, melicher2018riding}.

For reflected XSS, we inject generated payloads into URLs and test whether they execute. For stored XSS, we load the page, write the payload into browser storage, reload, and observe whether execution occurs, indicating insertion into an unsafe sink such as \texttt{innerHTML} or \texttt{eval} without sanitization. We record all confirmed vulnerabilities together with their exploitable taint flows.

\subsubsection{Security Headers Analysis.}
To analyze security-header configurations at scale, we instrumented our Playwright-based crawler with a network-event listener that records all request and response headers during page loads. For each domain, we consider only same-origin, document-type responses with status code 200 so that evaluations reflect the domain’s own configuration rather than that of third-party or redirected hosts.

For every qualifying response, we extract security-relevant headers and check them against predefined correctness criteria. A domain is marked as misconfigured if any valid page violates these rules. We analyze six classes of security-sensitive headers: content injection protection, clickjacking protection, cookie security, HSTS, CORS, and content-type/MIME handling. Detailed definitions are given in \Cref{sec:header}.

\subsubsection{TLS Misconfiguration Analysis.}
To identify TLS misconfigurations, we use Foxhound’s error logs, which capture connection failures and certificate-validation errors during page loads. For each domain, we load its landing page in Foxhound and record any TLS errors raised during the handshake or certificate checks. We focus on six common categories: domain mismatch, untrusted certificate authority, expired certificates, revoked certificates, unsupported protocol versions, and disabled signature algorithms. If Foxhound reports any such error, we flag the domain as TLS-misconfigured. This yields a consistent, browser-validated view of TLS correctness without requiring active probing beyond a standard page load.

\begin{table*}
\centering
\small
\setlength{\tabcolsep}{1.5pt}
\begin{adjustbox}{max width=\textwidth}
\begin{tabular}{l  rr rr rr rr rr rr rr rr rr}
\toprule
 &
\multicolumn{2}{c}{\textbf{Content type}} &
\multicolumn{2}{c}{\textbf{Clickjacking}} &
\multicolumn{2}{c}{\textbf{CORS}} &
\multicolumn{2}{c}{\textbf{Cookie}} &
\multicolumn{2}{c}{\textbf{HSTS}} &
\multicolumn{2}{c}{\textbf{Injection}} &
\multicolumn{2}{c}{\textbf{CXSS}} &
\multicolumn{2}{c}{\textbf{TLS}} &
\multicolumn{2}{c}{\textbf{Total}} \\
\cmidrule(lr){2-3}
\cmidrule(lr){4-5}
\cmidrule(lr){6-7}
\cmidrule(lr){8-9}
\cmidrule(lr){10-11}
\cmidrule(lr){12-13}
\cmidrule(lr){14-15}
\cmidrule(lr){16-17}
\cmidrule(lr){18-19}
\textbf{Tranco Ranking}
& Count & \%
& Count & \%
& Count & \%
& Count & \%
& Count & \%
& Count & \%
& Count & \%
& Count & \%
& Count & \% \\
\midrule

\hspace{1em}1--10K &
74   & 0.74\% &
1,184 & 11.84\% &
96   & 0.96\% &
2,335 & 23.35\% &
1,383 & 13.83\% &
2,849 & 28.49\% &
20   & 0.20\% &
47   & 0.47\% &
3,926 & 39.26\% \\

\hspace{1em}10,001--50K &
201  & 0.50\% &
4,401 & 11.00\% &
291  & 0.73\% &
9,144 & 22.86\% &
5,624 & 14.06\% &
10,906& 27.27\% &
79   & 0.20\% &
263  & 0.66\% &
15,966& 39.92\% \\

\hspace{1em}50,001--100K &
240  & 0.48\% &
5,420 & 10.84\% &
372  & 0.74\% &
11,863& 23.73\% &
7,324 & 14.65\% &
13,702& 27.40\% &
84   & 0.17\% &
485  & 0.97\% &
20,778& 41.56\% \\

\hspace{1em}100,001--200K &
424  & 0.42\% &
10,556& 10.56\% &
700  & 0.70\% &
25,768& 25.77\% &
15,900& 15.90\% &
26,654& 26.65\% &
162  & 0.16\% &
759  & 0.76\% &
42,662& 42.66\% \\

\hspace{1em}200,001--350K &
576  & 0.38\% &
13,825& 9.22\% &
749  & 0.50\% &
35,769& 23.85\% &
22,005& 14.67\% &
35,009& 23.34\% &
226  & 0.15\% &
1,372 & 0.91\% &
58,568& 39.05\% \\

\hspace{1em}350,001--500K &
680  & 0.45\% &
13,077& 8.72\% &
686  & 0.46\% &
31,469& 20.98\% &
20,529& 13.69\% &
30,826& 20.55\% &
235  & 0.16\% &
1,494 & 1.00\% &
52,894& 35.26\% \\

\midrule
\textbf{C. Crawl Tot.} & 4,079,855 & 16.43\% & 4,588,855 & 18.48\% & 69,546 & 0.28\% & 7,767,521 & 31.28\% & 6,710,461 & 27.02\% & 3,647,031 & 14.69\% & n.a. & n.a. & n.a. & n.a. & 13,594,429 & 54.74\% \\

\bottomrule
\end{tabular}
\end{adjustbox}
\caption{Vulnerability Metrics by Domain Ranking Range}
\label{tab:vuln_range}
\end{table*}

In Table~\ref{tab:vuln_range}, we present an overview of the number of domains affected by each security issue in our two datasets. For Tranco, domains are grouped by ranking range to show how issue counts vary across popularity levels. For Common Crawl, we report the corresponding overall counts on our dataset.

\section{Assessing Impact}
\label{sec:sampleeval}
\label{sec:impact}

\begin{figure*}
    \centering
    \includegraphics[width=\linewidth]{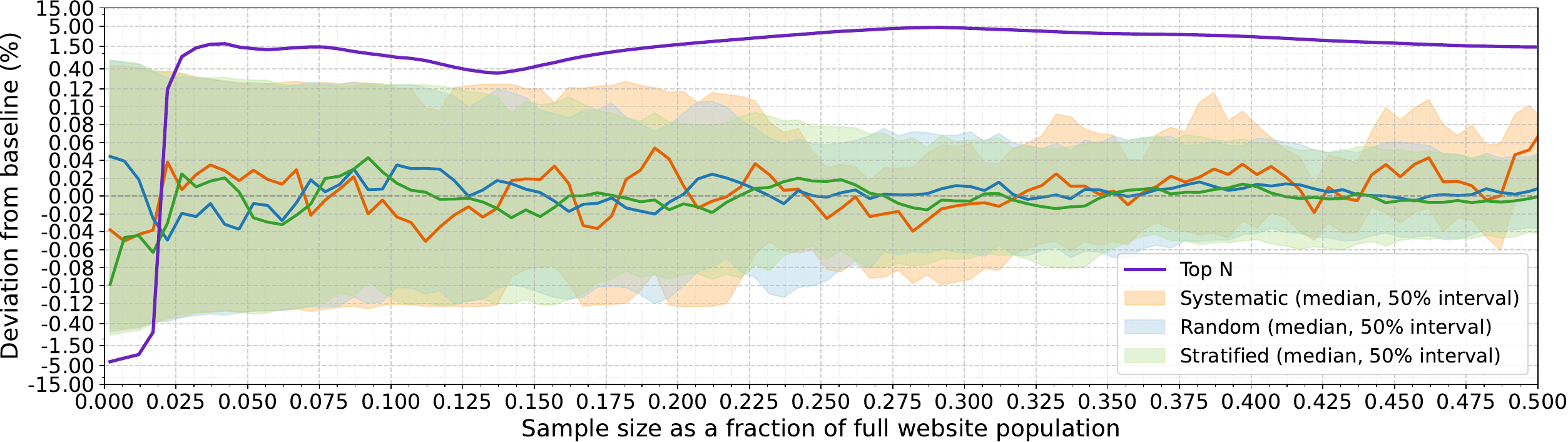}
    \caption{Deviation from the true impact across increasing sampling ratios for the full 500k baseline (5-window smoothing; Y axis on log scale).}
    \label{fig:deviation_vs_sample_ratio_smooth}
\end{figure*}

\begin{table}
\centering
\small
\setlength{\tabcolsep}{3pt}
\begin{tabular}{l | rrr | rr | rrr}
\toprule
&
\multicolumn{3}{c|}{\textbf{Stability}} &
\multicolumn{2}{c|}{\textbf{Run-wise}} &
\multicolumn{3}{c}{\textbf{Majority}}
\\
Sampling& 
ZCR & 
Peak & 
RMS & 
Over- & 
Under- & 
Over- & 
Under- & 
Tie \\
\midrule
\rowcolor{gray!15}\multicolumn{9}{l}{\textbf{Baseline = 500k}} \\
Top N        & 1  & 8.659 & 2.619 & 95.0\% &  5.0\% & - &  - & -\\
Systematic   & 47 & 2.841 & 0.234 & 50.6\% & 49.4\% & 51.5\% & 42.6\% & 5.9\%\\
Random       & 53 & 3.159 & 0.242 & 51.5\% & 48.5\% & 55.4\% & 42.6\% & 2.0\%\\
Stratified   & 48 & 6.741 & 0.270 & 49.8\% & 50.2\% & 41.6\% & 48.5\% & 9.9\%\\
\bottomrule
\end{tabular}
\caption{Stability and approximation analysis for impact estimation on the full 500k baseline.}
\label{tab:aggregated}
\end{table}

We evaluate how sampling strategies estimate \emph{impact}, defined here as the fraction of vulnerable domains within a chosen Tranco baseline. For each strategy, we compare the estimate obtained from the sample against the corresponding baseline value, which we refer to as the \emph{true impact}. We study three questions: how the basic sampling strategies behave on the full list and on Top \(K\) prefixes, how hybrid strategies behave, and how probability sampling changes across vulnerability classes with different prevalence.

We measure \emph{accuracy} as deviation from the true impact. For randomized strategies, each sampling ratio is evaluated over 50 independent runs. In the figures, we report the median together with the central 50\% interval. We use this interval for readability: wider intervals substantially obscure comparisons between overlapping strategies. We do not use it to characterize worst-case behavior. Instead, worst-case spread is captured separately through the Peak metric, which reports the maximum observed deviation across runs. Implementation details for the individual samplers, including our treatment of systematic and stratified sampling, are provided in \Cref{sec:appendix_sampling_strategies}.

We measure stability by analyzing how the error evolves across sampling using: (i) the Zero Crossing Rate (ZCR), which quantifies how often a method switches between overestimation and underestimation where a zero crossing occurs when the signed error changes from positive to negative, or vice versa, and a higher ZCR indicates more frequent oscillation around the baseline; (ii) Peak deviation, which captures the worst-case drift from the baseline; and (iii) Root Mean Square (RMS) error, which indicates the typical magnitude of deviation with greater emphasis on larger errors. Run-wise Over/Under: We summarize all individual estimates across all sample sizes and replicate runs and calculate the proportions above and below the baseline. Majority Over/Under/Tie: For each sample size, we compare the numbers of overestimations and underestimations across 50 runs and then count how many sample sizes have a majority of overestimations, a majority of underestimations, or a tie.

\subsection{Basic Sampling Methods}

\subsubsection{Full-list Baseline (500k)}
\label{sec:basic_impact}

\Cref{fig:deviation_vs_sample_ratio_smooth} and \Cref{tab:aggregated} show a clear split between Top \(N\) and probability-based sampling. Top \(N\) remains far from the true impact across the sampling range, whereas Random, Systematic, and Stratified sampling stay tightly centered around it.

\paragraph{Top N.}
Top \(N\) is persistently biased as an estimator of the full-list baseline. Its deviation reaches nearly \(\pm 9\) percentage points at small sampling ratios and remains large as the sample grows. This is reflected in its peak deviation (8.659), RMS (2.619), and near-zero crossing behavior (ZCR \(=1\)). The direction of error is also highly asymmetric: Top \(N\) over-estimates the true impact in 95\% of runs. The apparent smoothness of its curve therefore does not indicate stability in the statistical sense; it reflects a consistent structural bias induced by the ranking prefix.

\paragraph{Probability Sampling.} {Random, Systematic, and Stratified sampling behave differently. Their deviations remain small, oscillate around the true impact. Also, deviations of random and stratified sampling narrow as the sample grows. Importantly, we find that the RMS of the probabilistic sampling strategies are below 0.3, way lower compared to the Top N sampling of 2.6. The peak deviation is also lower compared to Top N sampling. }

\paragraph{Takeaway.}

{For the full 500k baseline, the main result is not subtle: Top \(N\) is biased, while probability-based strategies are accurate and stable. Systematic sampling provides the strongest overall performance in this setting.}

\subsubsection{Top \(K\) Baselines (10k, 20k, 50k, 100k)}

The full 500k baseline captures impact over the entire ranked population, but many prior studies do not sample from the full ranking. Instead, they first restrict the sampling universe to a prefix and then draw samples within that prefix. For example, Lee et al.~\cite{lee2023adcpg} evaluated their proposed approach on a 10k random sample from the Top 100K. {We model this design choice by repeating the analysis on Top 10K, 20k, 50k and 100K baselines. For each prefix, we recompute the true impact within that baseline and evaluate the samplers against that value, with the minimal sample size of 5\% of the target baseline sample size. {We exclude Stratified sampling here, as stratification is designed for broader ranked populations; on narrow prefixes, proportional sampling from deeper strata would not match the measurement objective.}}

\begin{figure*}
    \centering
    \begin{subfigure}{0.49\textwidth}
        \centering
        \includegraphics[width=\textwidth]{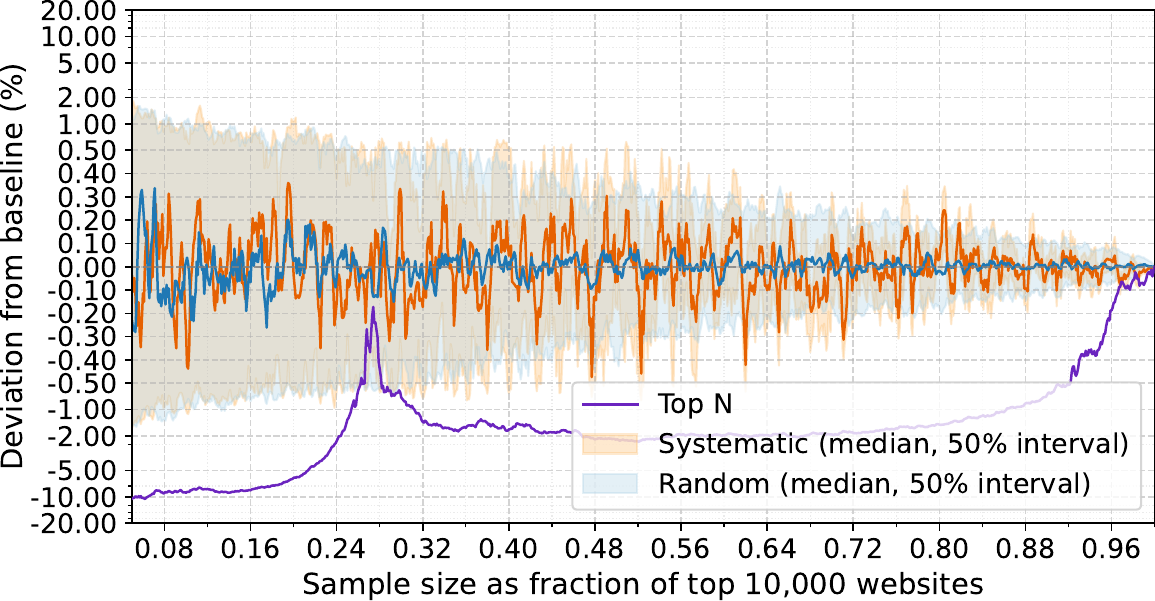}
        \caption{Top 10K baseline}
        \label{fig:impact10k_vs_sample_ratio_smooth}
    \end{subfigure}
    \hfill
    \begin{subfigure}{0.49\textwidth}
        \centering
        \includegraphics[width=\textwidth]{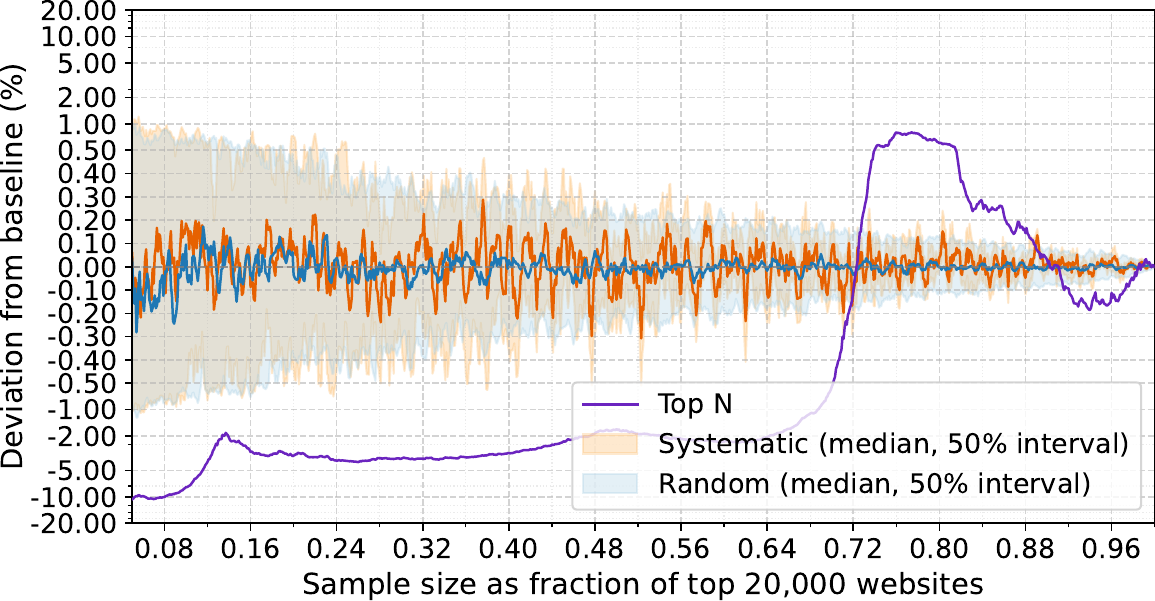}
        \caption{Top 20K baseline}
        \label{fig:impact20k_vs_sample_ratio_smooth}
    \end{subfigure}
    
    \begin{subfigure}{0.49\textwidth}
        \centering
        \includegraphics[width=\textwidth]{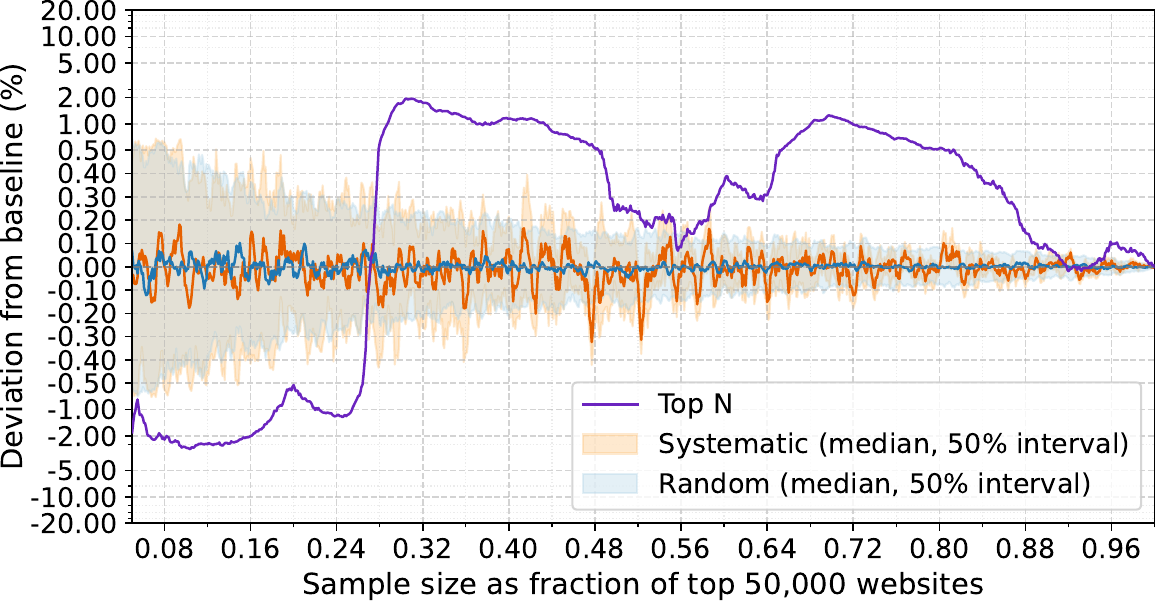}
        \caption{Top 50K baseline}
        \label{fig:impact50k_vs_sample_ratio_smooth}
    \end{subfigure}
    \hfill
    \begin{subfigure}{0.49\textwidth}
        \centering
        \includegraphics[width=\textwidth]{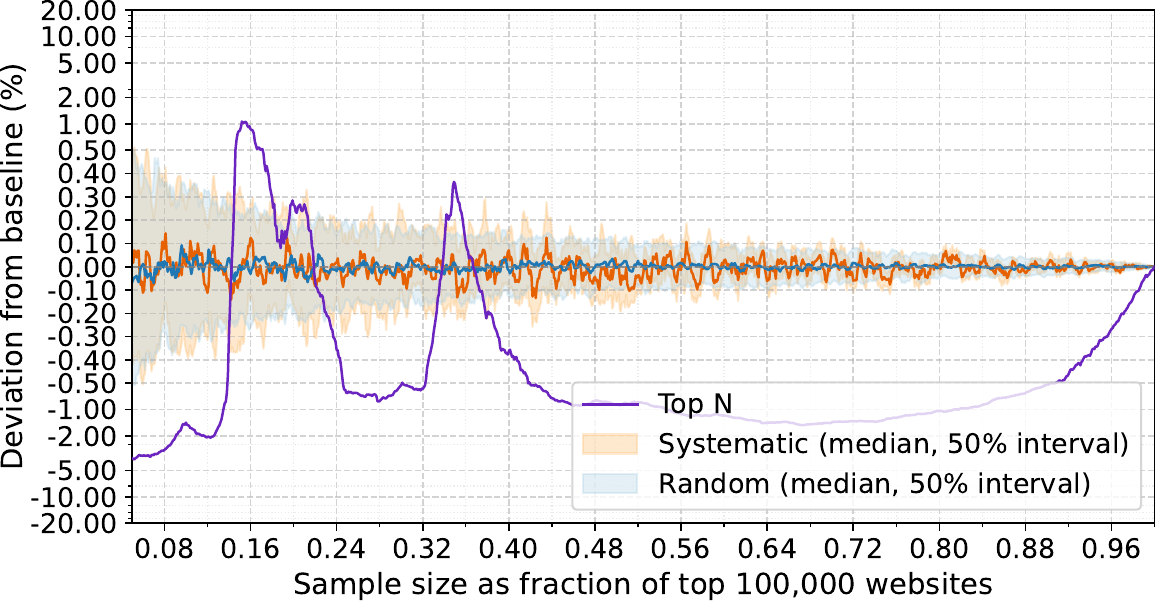}
        \caption{Top 100K baseline}
        \label{fig:impact100k_vs_sample_ratio_smooth}
    \end{subfigure}
    
    \caption{Deviation from the true impact across increasing sampling ratios for Top \(K\) baselines (5-window smoothing; Y axis on log scale).}
    \label{fig:impact_deviation_comparison}
\end{figure*}

\begin{table}[t]
\centering
\small
\setlength{\tabcolsep}{3pt}
\begin{tabular}{l | rrr | rr | rrr}
\toprule
&
\multicolumn{3}{c|}{\textbf{Stability}} &
\multicolumn{2}{c|}{\textbf{Run-wise}} &
\multicolumn{3}{c}{\textbf{Majority}}
\\
\textbf{Sampling} &
ZCR &
Peak &
RMS &
Over- &
Under- &
Over- &
Under- &
Tie \\
\midrule
\rowcolor{gray!15}\multicolumn{9}{l}{\textbf{Baseline = 10k}} \\
Top N        & 0   & 10.550 & 3.767 & 0  & 100 & - &  - & -\\
Random       & 461 & 8.317  & 0.724 & 50.2 & 49.8 & 45.3 & 43.7 & 11.0 \\
Systematic   & 491 & 6.145  & 0.705 & 48.9 & 50.2 & 46.3 & 50.1 & 3.6 \\
\midrule
\rowcolor{gray!15}\multicolumn{9}{l}{\textbf{Baseline = 20k}} \\
Top N        & 9   & 10.635 & 3.269 & 20.5  & 79.5 & - &  - & -\\
Random       & 463 & 5.233  & 0.514 & 49.6 & 50.4 & 43.0 & 48.5 & 8.5 \\
Systematic   & 486 & 4.003  & 0.497 & 51.0 & 49.0 & 50.2 & 47.7 & 2.1 \\
\midrule
\rowcolor{gray!15}\multicolumn{9}{l}{\textbf{Baseline = 50k}} \\
Top N        & 5   & 2.842 & 1.088 & 74.5  & 25.5 & - &  - & -\\
Random       & 482 & 3.031  & 0.323 & 50.0 & 50.0 & 43.8 & 42.8 & 13.4 \\
Systematic   & 507 & 2.698  & 0.333 & 48.8 & 51.2 & 46.0 & 51.0 & 3.0 \\
\midrule
\rowcolor{gray!15}\multicolumn{9}{l}{\textbf{Baseline = 100k}} \\
Top N        & 4   & 3.728 & 1.120 & 10.9 & 89.1 & - &  - & -\\
Random       & 478 & 2.420  & 0.228 & 49.9 & 50.1 & 43.6 & 46.2 & 10.2 \\
Systematic   & 501 & 2.199  & 0.234 & 49.0 & 51.0 & 45.1 & 50.3 & 4.6 \\
\bottomrule
\end{tabular}
\caption{Stability and approximation analysis for impact estimation on Top \(K\) baselines.}
\label{tab:maxsize}
\end{table}

{\Cref{fig:impact_deviation_comparison} and \Cref{tab:maxsize} show the same qualitative pattern as in the full-list experiment, but with higher early variance. On smaller prefixes, vulnerable domains are sparser in absolute terms, so early samples are more sensitive to whether they include one of the few positives. This increases volatility at low sampling ratios especially for Top 10K and Top 20K, and the perturbation range in deviation narrows as baseline size grows from 10K to 100K. However, it does not change the overall character of the estimators.}

\paragraph{Probability Sampling.}
{Random and Systematic sampling remain centered around the true impact across all Top \(K\) baselines. Their zero-crossing rates are high because the estimates oscillate frequently around the baseline, not because they drift far from it. The more informative quantities here are Peak and RMS: both methods improve steadily as the baseline grows, with Random achieving the lower RMS in every setting. By Top 100K, both estimators are already tightly concentrated, with RMS values of 0.228 and 0.234, respectively.}

\paragraph{Top N.}
Top \(N\) again behaves as a biased estimator rather than a noisy one. Its errors remain large across all prefixes, but the sign depends on the baseline: it under-estimates for some prefixes and over-estimates for others. This shift in direction underscores the core problem with Top \(N\): its error is driven by the structure of the ranking prefix, not by ordinary sampling variance. {Moreover, Top \(N\) demonstrates significantly low ZCR close to 0 and largely imbalanced distribution, demonstrating an unstable performance. This stability issue has also contributed to the peak deviation being slightly below the Random sampling by 0.2\%, only for the 50k baseline (2.84\% vs. 3.03\%). However, the RMS of top \(N\) is thrice as big as that of Random sampling, with 74.5\% run-wise over-estimation versus fully balanced performance of Random sampling, indicating that overall bias is still in effect. }

\paragraph{Takeaway.}
Across top \(K\) baselines, probability sampling remains reliable even when early estimates are noisy. The extra volatility at small sampling ratios is a consequence of sparse positives, not estimator bias. Top \(N\), in contrast, continues to produce large structural errors as an estimator of the full-baseline.

\subsection{Hybrid Sampling}

Hybrid samplers combine a deterministic Top \(N\) prefix with a probability-based sample drawn from the remainder of the ranking. The appeal is straightforward: include prominent domains by construction, then broaden coverage through randomization. We evaluate both whether this improves impact estimation and which component drives the final result.

\subsubsection{Impact Estimation}
\label{sec:res_hybrid_impact}

\begin{figure}
    \centering
    \includegraphics[width=\linewidth]{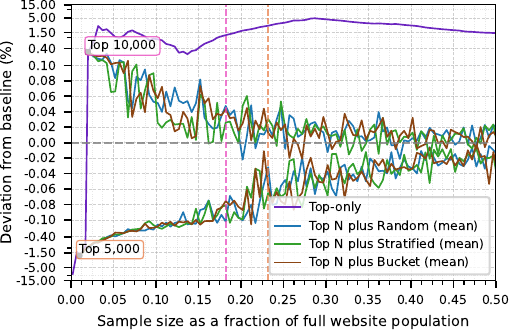}
    \caption{Deviation from the true impact for hybrid sampling strategies across increasing sampling ratios (5-window smoothing; Y axis on log scale).}
    \label{fig:hybrid_deviation_vs_sample_ratio_smooth}
\end{figure}

\begin{table}[t]
\centering
\small
\setlength{\tabcolsep}{3pt}
\begin{tabular}{l | rrr | rr | rrr}
\toprule
&
\multicolumn{3}{c|}{\textbf{Stability}} &
\multicolumn{2}{c|}{\textbf{Run-wise}} &
\multicolumn{3}{c}{\textbf{Majority}}
\\
Sampling &
ZCR &
Peak &
RMS &
Over- &
Under- &
Over- &
Under- &
Tie \\
\midrule
\rowcolor{gray!15}\multicolumn{9}{l}{\textbf{Top N = 5k}} \\
Random     & 0 & 2.025 & 0.279 & 26.7 & 73.3 & 0   & 100 & 0 \\
Stratified & 2 & 2.132 & 0.278 & 27.5 & 72.5 & 1   & 96.9 & 2.1 \\
Bucket     & 0 & 2.012 & 0.277 & 26.9 & 73.1 & 0   & 100 & 0 \\
\midrule
\rowcolor{gray!15}\multicolumn{9}{l}{\textbf{Top N = 10k}} \\
Random      & 16 & 0.858 & 0.142 & 57.9 & 42.1 & 78.1 & 10.4 & 11.5 \\
Stratified  & 18 & 1.117 & 0.147 & 58.1 & 41.9 & 78.1 & 14.6 & 7.3 \\
Bucket      & 19 & 0.790 & 0.147 & 57.4 & 42.6 & 76.0 & 17.7 & 6.3 \\
\bottomrule
\end{tabular}
\caption{Stability and approximation analysis for hybrid Top \(N\) sampling strategies.}
\label{tab:hybrid_topn}
\end{table}

\Cref{fig:hybrid_deviation_vs_sample_ratio_smooth} shows that hybrid estimators inherit the bias of their Top \(N\) prefix and spend the remainder of the sample budget correcting it. The initial jump at the Top \(N\) cutoff reproduces the standalone behavior of Top \(N\): large for Top 5K, smaller for Top 10K, but non-negligible in both cases. Beyond that cutoff, the hybrid curves begin to resemble the corresponding probability samplers.

This correction is real but costly. For Top 10K hybrids, the probabilistic tail can eventually pull the estimate close to the true impact, as reflected in the low RMS values in \Cref{tab:hybrid_topn}. For Top 5K hybrids, the inherited bias is much larger and remains visible much longer. In other words, hybrids do improve as the tail grows, but the early sample budget is spent offsetting the deterministic prefix rather than improving over a pure probability sample.

The three tail samplers behave similarly at the aggregate level. Random, Stratified, and Bucket produce near-identical {with less than 0.3\% difference of} peak errors for a fixed prefix, and their RMS values differ only marginally {by 0.005\%}. {However, same sampling strategy differs drastically across different Top N prefixes by up to 100\% in terms of peak deviation and RMS.} The dominant factor is therefore the prefix size, not the choice of tail sampler.

\paragraph{Takeaway.}
Hybrid samplers do not escape the central weakness of Top \(N\). Their performance is largely determined by how much bias the deterministic prefix injects and how much probabilistic tail sampling is required to undo it.

\subsubsection{Contribution Analysis via Shapley Values}
\label{sec:res_hybrid_shapley}

The hybrid curves show that the tail corrects the prefix. We now ask \emph{which} component contributes error and which contributes recovery. To answer this, we use Shapley values~\cite{winter2002shapley} to decompose the hybrid estimator into two components: the Top \(N\) block (\(A\)) and the rest-sampler (\(B\)).

We fix the prefix at Top 5K and measure its standalone error. For each rest-sampling mode (Random, Bucket, Stratified), we then draw rest-samples from 1,000 to 50,000 domains, repeating each draw 50 times. For every rest-sample size, we compute the error of the Top \(N\) block alone (\(E_A\)), the error of the rest-sample alone (\(E_B\)), and the error of the combined hybrid (\(E_{AB}\)). A detailed derivation is provided in \Cref{sec:shapley_detailed}.

\begin{figure}
    \centering
    \includegraphics[width=\linewidth]{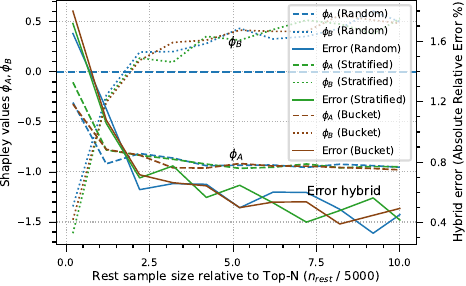}
    \caption{Shapley values of the Top \(N\) block (\(\phi_A\)) and the rest-sampling component (\(\phi_B\)) as the rest-sample grows from 1k to 50k domains (0.2\(\times\) to 10\(\times\) the Top \(N\) size).}
    \label{fig:shapley_hybrid}
\end{figure}

\Cref{fig:shapley_hybrid} makes the division of labor explicit. Across all three hybrid strategies, the Top \(N\) block has strictly negative Shapley values throughout. Its contribution is therefore not merely imperfect; it is systematically harmful. The deterministic prefix starts the estimator in a biased state and continues to reduce accuracy even as the rest-sample grows.

The rest-sampling component behaves differently. At very small sizes, it also contributes negatively because its own variance is high. Once the tail becomes large enough, however, its Shapley value turns positive and remains so. This is the point at which the probabilistic component begins to reduce total hybrid error rather than add to it. Stratified and Bucket sampling cross this threshold slightly earlier than Random, consistent with their lower variance.

\paragraph{Takeaway.}
The Shapley decomposition confirms the interpretation suggested by the hybrid curves: hybrid accuracy comes from the probabilistic tail, not from the deterministic prefix. The Top \(N\) block contributes negative value throughout; the rest-sampler is the only component that eventually improves the estimate.

\subsection{Sensitivity Across Vulnerability Buckets}

Our analyses so far aggregate all security issues into a single vulnerable population. This reveals overall sampling behavior, but it does not show whether the same pattern holds for issues with very different prevalence. We therefore group vulnerabilities by the number of affected domains and re-evaluate probability sampling within each group.

We restrict this analysis to Random, Systematic, and Stratified sampling. The point here is to test how estimator quality changes as prevalence changes. Since Top \(N\) is a deterministic prefix selector rather than an unbiased estimator of the full baseline, it is not informative for this comparison.

\paragraph{Buckets.}
\Cref{tab:vuln_buckets} groups the issues in our dataset by the number of affected domains, from rare classes with only a few hundred positives to common classes affecting well over 100k domains.

\begin{table}[t]
\centering
\small
\begin{tabular}{l l}
\toprule
\textbf{\# Domains} & \textbf{Included Issues} \\
\midrule
0--1{,}000           & Client-side XSS (CXSS) \\
2{,}000--3{,}000     & Content-Type \& MIME Handling, CORS \\
4{,}000--5{,}000     & TLS misconfiguration \\
40{,}000--50{,}000   & Clickjacking (missing or misconfigured) \\
70{,}000--80{,}000   & HSTS (missing or misconfigured) \\
110{,}000--120{,}000 & Content Injection, Cookie Security \\
\bottomrule
\end{tabular}
\caption{Buckets of the robustness experiment, grouped by the number of vulnerable domains.}
\label{tab:vuln_buckets}
\end{table}

For each bucket, we compute how often the 150 estimates at a given sampling ratio (50 runs for each of the three probability samplers) fall within a 5\% relative-error tolerance of the true impact.

\begin{figure}[t]
    \centering
    \includegraphics[width=\columnwidth]{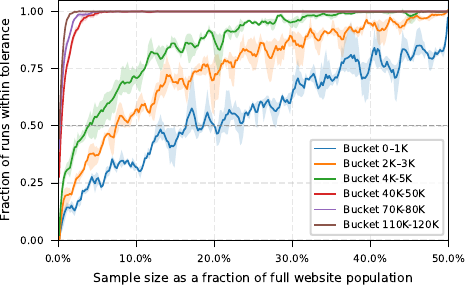}
    \caption{Fraction of probability-sampling runs within a 5\% relative-error tolerance of the true impact across vulnerability buckets of different sizes (11-window smoothing).}
    \label{fig:bucket_robustness}
\end{figure}

\begin{figure*}
    \centering
    
    \begin{subfigure}[b]{0.49\linewidth}
        \centering
        \includegraphics[width=\linewidth]{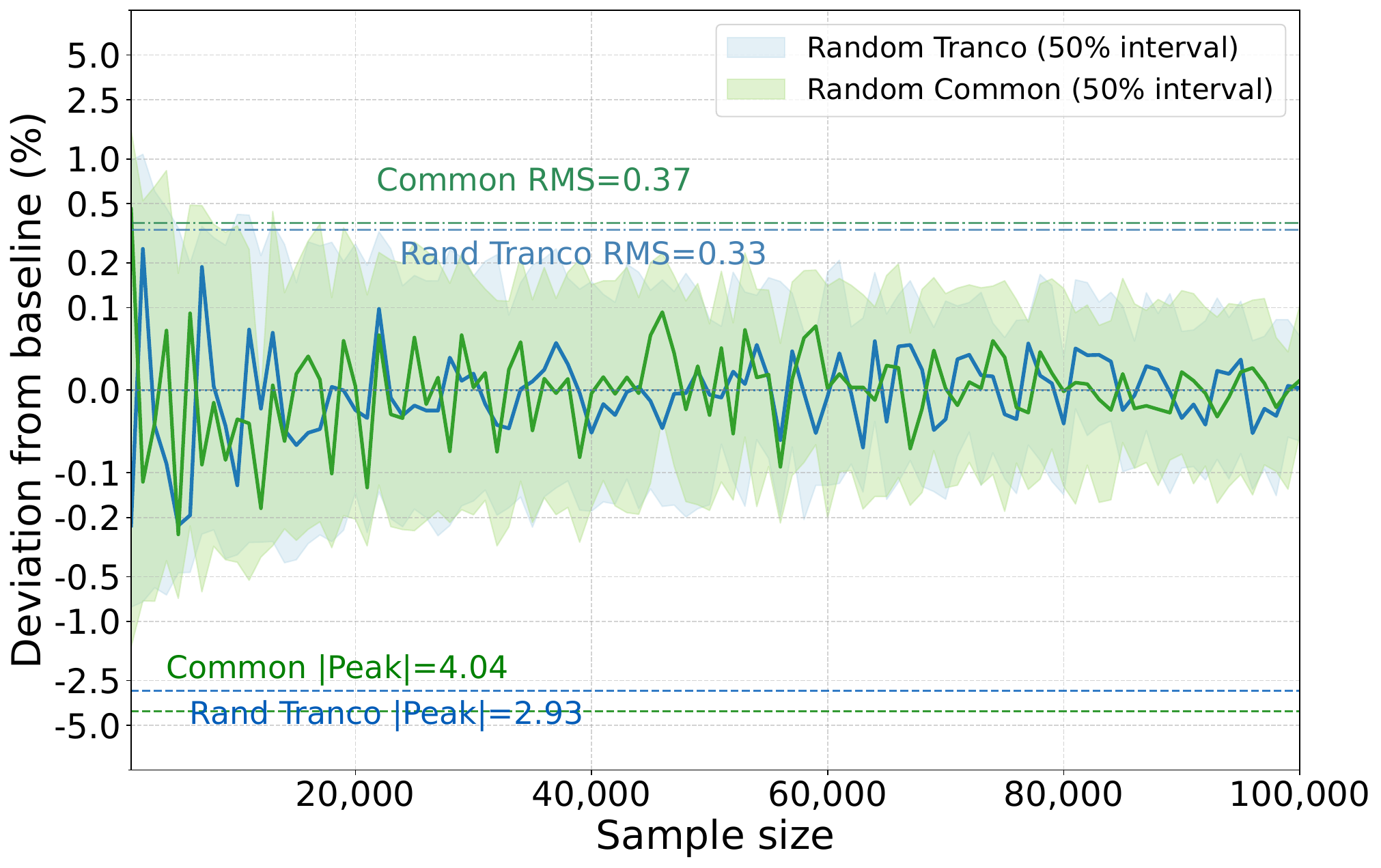}
        \Description{A plot showing the deviation from prevalence for the Tranco baseline.}
        \caption{Respective Baselines}
    \end{subfigure}\hfill
    \begin{subfigure}[b]{0.49\linewidth}
        \centering
        \includegraphics[width=\linewidth]{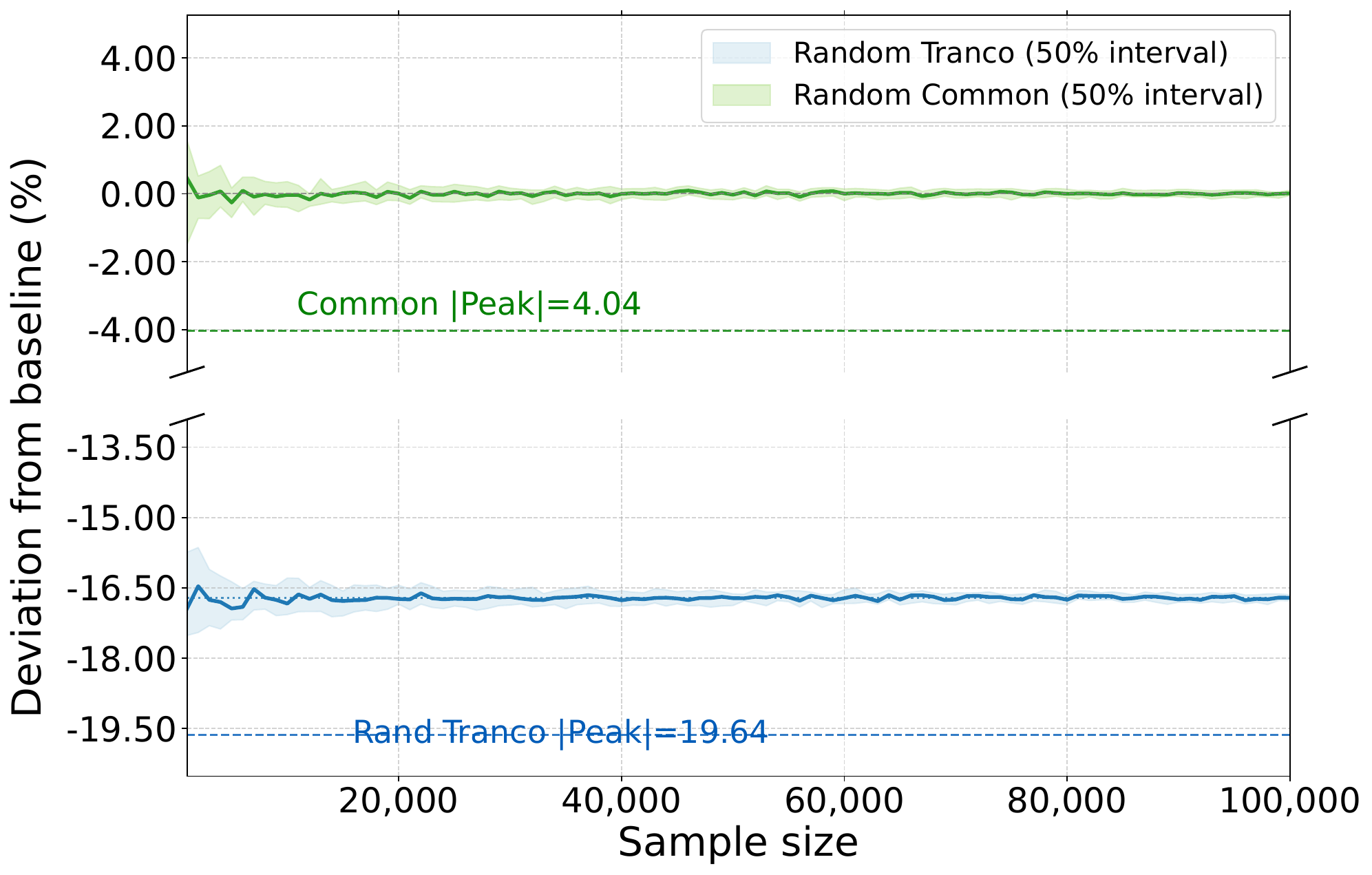}
        \Description{A plot showing the deviation from prevalence for the Common Crawl baseline.}
        \caption{Common Crawl baseline}
    \end{subfigure}
    \caption{Deviation of Sample Result w.r.t. Common Crawl Baseline and Self Baseline. Fig. a) shows the relative deviation with respect to the datasets' own ground truth. Fig. b) shows the deviation with respect to Common Crawl ground truth.}
    \label{fig:prevalence}
\end{figure*}

\paragraph{Takeaway.}
\Cref{fig:bucket_robustness} shows a consistent prevalence-driven pattern. For common issues, probability sampling reaches the true impact quickly and with little uncertainty. For rarer issues, convergence is slower and larger samples are required, but the curves still improve smoothly as the sample grows. The main implication is that probability sampling remains well-behaved across vulnerability classes; what changes is not the qualitative behavior of the estimator, but the amount of data needed for it to stabilize.

\begin{table}
\centering
\small
\setlength{\tabcolsep}{2pt}
\begin{tabular}{l | rrr | rr | rrr}
\toprule
&
\multicolumn{3}{c|}{\textbf{Stability}} &
\multicolumn{2}{c|}{\textbf{Run-wise}} &
\multicolumn{3}{c}{\textbf{Majority}}
\\
Dataset & 
ZCR & 
Peak & 
RMS & 
Over- & 
Under- & 
Over- & 
Under- & 
Tie \\
\midrule
Tranco  & 41 & 2.93 & 0.33 & 49.9\% & 50.1\% & 38\% & 48\% & 14\%\\
Common Crawl  & 56 & 4.04 & 0.37 & 49.7\% & 50.3\% & 48\% & 39\% & 13\%\\
\bottomrule
\end{tabular}
\caption{Merged results for the prevalence analysis for \Cref{fig:prevalence}a.}
\label{tab:merged_commconcrawl_for_a}
\end{table}

\begin{figure}
    \centering
    \includegraphics[width=\columnwidth]{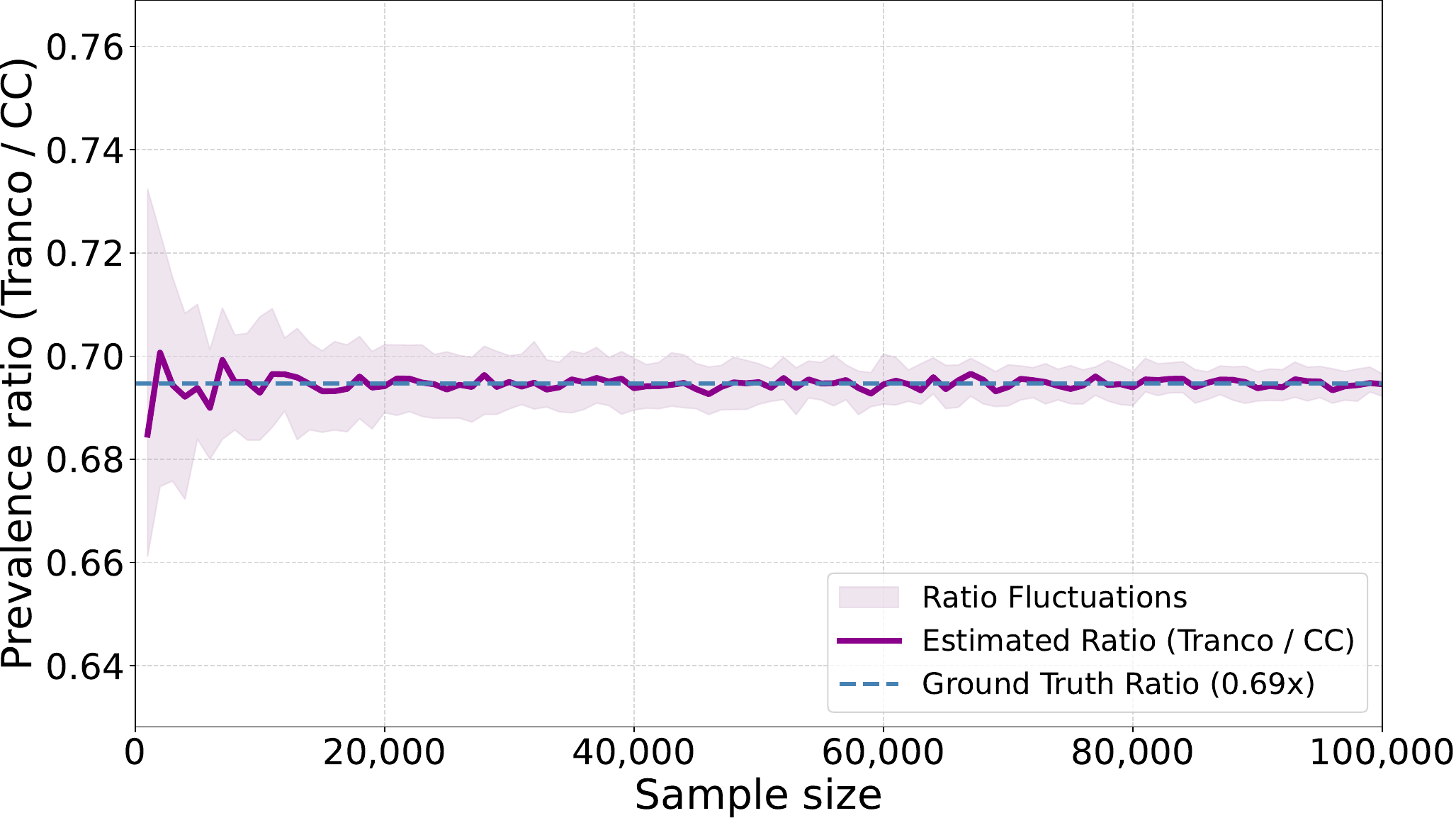}
    \caption{Prevalence estimation ratio between Common Crawl and Tranco across sample sizes}
    \label{fig:prevalence_ratio}
\end{figure}

\section{Prevalence}
\label{sec:prevalence}

Prevalence measurements aim to estimate how frequently a security issue occurs on the Web at large. This objective differs from impact estimation, where the focus is on the fraction of vulnerable domains within a selected popularity-ranked population. For prevalence, the relevant question is not how many domains in a chosen list are vulnerable, but what fraction of domains in the broader web population is vulnerable. We therefore use Common Crawl as the reference population and define the baseline prevalence as the fraction of vulnerable hosts in our processed Common Crawl dataset.

\subsection{Setup and Dataset Scope}
\label{sec:prevalence-setup}

Because Common Crawl is not a popularity-ranked list like Tranco, rank-dependent strategies such as Top $N$ are not applicable in this setting. For the same reason, Systematic and Stratified sampling are not meaningful here, as they require a meaningful ordering or rank structure. We therefore restrict the prevalence evaluation to Random sampling.

We exclude client-side XSS and TLS misconfigurations from this analysis. Common Crawl provides a static archival snapshot of the Web, and evaluating these two classes would require revisiting hundreds of millions of pages with dynamic instrumentation or TLS handshakes, which is infeasible at this scale. Still, the remaining classes already reveal substantial differences between the two datasets.

\subsection{Differences Between Tranco and Common Crawl}
\label{sec:prevalence-dataset-differences}

Before evaluating sampling behavior, we first compare the fraction of vulnerable domains in Tranco and Common Crawl. \Cref{tab:vuln_range} shows that the two populations differ markedly. Common Crawl has substantially higher prevalence for Content Type issues (16.43\% vs.\ 0.44\%), Clickjacking (18.48\% vs.\ 9.69\%), Cookie security (31.28\% vs.\ 23.27\%), and HSTS (27.02\% vs.\ 14.55\%). Tranco exceeds Common Crawl only for CORS (0.58\% vs.\ 0.28\%) and Injection (23.99\% vs.\ 14.69\%).

These differences matter because they show that popular domains are not representative of the broader host population observed in Common Crawl for prevalence estimation. The mismatch is not a uniform offset: for some issue classes the prevalence gap is large, and for others it even changes direction. As a result, sampling from Tranco may distort both the absolute prevalence estimate and the relative profile of which issues appear most widespread.

\subsection{Random Sampling for Prevalence Estimation}
\label{sec:prevalence-random}

\Cref{fig:prevalence} evaluates Random sampling under the Common Crawl prevalence baseline.  In this figure, we evaluate the performance of random sampling from both Tranco (blue regions) and Common Crawl (green regions). Similarly to previous figures, we plot the deviation from the baseline with median and the central 50\% interval. The two subplots utilize different baselines. In \Cref{fig:prevalence}a, we plot the deviation for both datasets with their datasets' own baseline to reflect the stability of the sampling strategy itself. In \Cref{fig:prevalence}b, we plot the deviation for both dataset with the baseline from Common Crawl to reflect the potential bias introduced by the dataset choice.

As shown in \Cref{fig:prevalence}a, random sampling based on both datasets performs tightly centered around zero deviation with rapid convergence when sampling size increases, which indicates that the random sampling strategy remains stable and approximately unbiased with respect to each dataset's own baseline. \Cref{tab:merged_commconcrawl_for_a} further shows detailed data for \Cref{fig:prevalence}a with a balanced run-wise over- and under-estimation ratio, which proves the stability of random sampling across datasets. However, despite per-dataset deviation consistency, \Cref{fig:prevalence}b demonstrates drastic deviation for random sampling {towards -16.5\%, with a peak absolute deviation of 19.6\%}. This gap does not close as sample size increases, and there is definitely no zero crossing.


To give a clear understanding of the ratio of underestimation, \Cref{fig:prevalence_ratio} visualizes the ratio of the median prevalence estimates from random sampling on Common Crawl to that from Tranco across sample sizes. When sample size increases, the ratio rapidly converges to 0.70$\times$. As such, relative to the broader Common Crawl baseline, random sampling from Tranco may underestimate the prevalence of the measured security issues by approximately 30\%, even with an appropriately large sample size.

All these results on the internal consistency of random sampling and drastic underestimation for mismatched dataset selection highlight the importance of dataset selection. If the dataset is not appropriate, merely optimizing sampling strategy and sizes cannot mitigate the inherited deviation of prevalence estimates.

\section{Practical Guidance for Sampling}
\label{sec:bestpractice}

Designing a web security measurement requires three decisions: which dataset
to sample from, which sampling strategy to apply, and how large a sample to
collect. The dataset determines the population that a measurement can
describe: a popularity-ranked list such as Tranco is natural for impact-oriented questions, whereas a broad archive such as Common Crawl better
matches prevalence-oriented questions, and the strategy determines how
faithfully a sample reflects that population. Our measurements show that both
choices have a first-order effect on the reported results: a deterministic
Top $N$ prefix can deviate from the rate of a larger population by several
percentage points; a probability sample drawn from the wrong population
remains far from the target even as the sample grows; and the sample size
needed for a reliable estimate depends strongly on the prevalence of the
issue. In the following, we first introduce an adaptive sampling strategy that
addresses the common situation in which the distribution of the studied issue
is unknown, evaluate it in a case study, and then summarize the resulting
guidance for impact- and prevalence-oriented measurements.

\subsection{Adaptive Probability Sampling}
\label{sec:adaptive}

Researchers often begin a study without knowing how common a vulnerability is or where affected domains are located in the list. A fixed sampling size is difficult to choose when the prevalence of the issue is unknown. If the size fraction is too small, the sample may not contain enough cases to reach the desired precision; if it is large enough for rare issues, it wastes resources on common ones. Our sensitivity analysis shows that the fraction required to achieve a given precision differs by orders of magnitude across prevalence levels, even though probability sampling itself behaves consistently. The sampling size should therefore be determined by the data rather than fixed in advance.

We therefore propose a new practical sampling strategy, adaptive probability
sampling, which begins with a 1--2\% probability-sampling pilot and expands
adaptively; most non-sparse issues stabilize by 5--10\% sampling. In practice,
the researcher can first draw a 2\% pilot sample and double the cumulative sample
through 4\%, 8\%, 16\%, etc., using nested simple random sampling without
replacement. At every later stage, the rule stops when the current estimate is positive and both the relative change from the preceding stage and the relative half-width of a high-confidence interval are sufficiently small; otherwise it doubles the sample, up to the complete dataset. The complete procedure and its confidence-interval details are given in 
 \Cref{sec:adaptive-sampling}.

\subsection{Case Study}

We evaluate the strategy on three target sample populations that cover the two measurement objectives: for impact, we measure two baselines, the Tranco Top 100K and Tranco Top 500K; and for the prevalence, we measure the strategy on our Common Crawl dataset. For each population, we select one high-, one medium-, and one low-rate outcome---Cookie security, Clickjacking, and CORS. We set the tolerance for both stopping conditions to 5\% and use a 95\% confidence level for the interval. Since the adaptive strategy is stochastic, we repeat each run 1{,}000 times for each population and outcome to check the overall results. We then compare the adaptive sampling results to the ground truth of the complete population and evaluate if the result falls within the 5\% relative tolerance  as accuracy.

\begin{table}[t]
\centering
\small
\setlength{\tabcolsep}{4pt}
\begin{tabular}{l rrrr}
\toprule
\textbf{Outcome} & \textbf{Median stop} & \textbf{Mean stop} & \textbf{Max stop} & \textbf{Accuracy} \\
\midrule
\rowcolor{gray!15}\multicolumn{5}{l}{\textbf{Tranco Top 100K}} \\
Cookie security & 8\%  & 8.14\%  & 16\%  & 99.0\% \\
Clickjacking    & 16\% & 16.37\% & 32\%  & 98.3\% \\
CORS            & 100\%& 100.00\%& 100\% & 100.0\% \\
\rowcolor{gray!15}\multicolumn{5}{l}{\textbf{Tranco Top 500K}} \\
Cookie security & 4\%  & 4.00\%  & 4\%   & 99.9\% \\
Clickjacking    & 4\%  & 4.10\%  & 8\%   & 99.2\% \\
CORS            & 64\% & 65.01\% & 100\% & 99.9\% \\
\rowcolor{gray!15}\multicolumn{5}{l}{\textbf{Common Crawl}} \\
Cookie security & 4\%  & 4.00\%  & 4\%   & 100.0\% \\
Clickjacking    & 4\%  & 4.00\%  & 4\%   & 100.0\% \\
CORS            & 4\%  & 4.02\%  & 8\%   &  99.0\% \\
\bottomrule
\end{tabular}
\caption{Adaptive sampling case study results over 1{,}000 retrospective trials. }
\label{tab:adaptive-case-studies}
\end{table}

In Table~\ref{tab:adaptive-case-studies} we show the results of our case
study. For each run, the strategy stops after sampling a certain fraction of
the population; we summarize these stopping fractions across the 1{,}000 runs
with three statistics: the median, i.e., the fraction at which half of the
runs stopped (the typical cost); the mean, i.e., the average stopping fraction
(the expected cost); and the maximum, i.e., the largest stopping fraction
among all runs (the worst-case cost). The accuracy reports the fraction of
runs whose stopped estimate is within 5\% relative error of the ground truth.

The results show that for common and moderately common issues, the median stopping fraction is only 4\%--16\% of the population, yet at least 98\% of the stopped estimates are within 5\% relative error of the full-population rate; in contrast, fixed 2\% and 4\% samples in Top 100K would have produced correct estimates for the sparse CORS outcome in only 10.4\% and 21.9\% of the runs, respectively. For rare issues, the strategy does not stop prematurely: in Top 100K it continues to the complete population, and in Top 500K it uses 64\% of the population, thereby making the cost of precision explicit instead of reporting an estimate with unsupported precision. The stopping decision reflects the information in the data rather than the sampling fraction alone: although CORS is also rarer in Common Crawl, a 4\% sample of the dataset can also produce satisfactory results that is because the Common Crawl is a massive dataset, 4\% already contains a substantial number of positive cases. The strategy therefore spends measurement budget only where the data require it, and it remains operational without prior knowledge of the issue distribution: the researcher only needs to declare the target population and the precision target.

\subsection{Recommendation}
The recommended practice consists of two elements, i.e., the choice of
dataset and the choice of sampling strategy, and heavily relies on the
research goal. In the following, we summarize our recommendations for the
sampling strategy of web security measurements.

\emph{1) Measuring impact.}
For impact-oriented measurements, the research question determines the choice
of sampling strategy. A popularity-based domain list such as Tranco is the
natural dataset, because it reflects the most popular websites. If the
researcher is only interested in the most popular sites, e.g., the state of
the top 1{,}000 domains, then Top $N$ is the obvious choice, and the results
should be explicitly reported as valid only for that prefix. In other cases,
the researcher may be interested in a much larger range of domains (e.g., the
top 500K domains), but cannot measure the entire population or merely wants
to conduct a small-scale study. Here, the research goal is to use minimal
samples to reflect the overall population of interest, so the researcher can
draw a probability sample and use the adaptive sampling strategy introduced
in Section~\ref{sec:adaptive} to properly reflect the distribution of the
large population of interest, and if resources allow only a small sample of websites, 
a plain probability sample remains safer than a Top $N$ prefix for this case.

\emph{2) Measuring prevalence.}
For prevalence, the goal is to approximate the rate of a security issue
across a large population. Contrary to impact measurement, comprehensiveness
matters more than popularity, and researchers should therefore choose a large
dataset such as Common Crawl rather than a smaller, ranked dataset such as
Tranco. Similarly, researchers can draw a probability sample from it using
the adaptive sampling strategy we propose.

\emph{3) Measuring both quantities.}
Since our prevalence measurements in \Cref{sec:prevalence} show substantially different rates between Tranco and Common Crawl, we recommend two explicitly separated analyses: an impact-oriented measurement over a popularity-ranked list such as Tranco, and a prevalence-oriented analysis over a broader archive such as Common Crawl. As sampling strategy and research question heavily affect the conclusion, we believe that it is better to separate the dataset for specific impact or prevalence objective and synthesize the conclusion.

\section{Discussion}
\label{sec:discussion}

\subsection{Lessons Learned}

We now distill the broader lessons from our study, focusing on the main methodological takeaways that generalize across sampling strategies, datasets, and measurement goals.

\emph{1) Sampling strategy should be chosen to match the measurement objective.}  
The measurement objective determines both the target dataset and the sampling
strategy: impact-oriented questions call for a ranked dataset such as Tranco,
while prevalence-oriented questions call for a broad archive such as Common
Crawl. Within the declared dataset, a complete measurement remains
appropriate when the target is exactly the Top $N$ prefix; otherwise, a
probability sample drawn from the whole dataset is recommended. We also
propose an adaptive sampling strategy for the common case where the
prevalence of the issue is unknown in advance. 

\emph{2) Default strategies (Top $N$) may not be the best choice.}
Our literature review shows that Top $N$ sampling is by far the dominant
strategy in prior web security measurements, and it is the natural choice when the measurement target is exactly the declared $N$ prefix. However, when it is used to estimate properties beyond that prefix, it introduces large, persistent bias, and this error does not disappear simply by increasing the sample size.

\emph{3) Probability sampling is more robust.}
Across all of our evaluations, probability-based methods can produce
small, stable, and approximately unbiased errors. On the full Tranco Top 500K
baseline, their RMS error remains below 0.3 percentage points, compared with
2.62 percentage points for Top $N$. Their estimates remain centered around
the baseline, their deviations are sharply bounded, and their accuracy and
stability improve predictably as the sample grows.

\emph{4) Hybrid strategies do not fix Top $N$ prefix, but inherit its bias.}  
Although hybrid methods appear to balance coverage of popular domains with
broader representativeness, in practice they spend a substantial part of
their sample budget correcting the bias introduced by their deterministic
Top $N$ prefix. Their performance is therefore determined largely by how
harmful that prefix is, rather than by any inherent advantage of the hybrid
design.

\emph{5) More data cannot fix the wrong population.}  
Sampling correctly from the wrong dataset still yields the wrong answer. In our prevalence analysis, Random sampling on Tranco is statistically well-behaved with respect to the Tranco population, but it remains systematically far from the Common Crawl baseline: its peak deviation from the Common Crawl baseline reaches {19.6 percentage points}. Increasing sample size reduces variance, but it does not eliminate the estimation error for prevalence.

\emph{6) Dataset choice must match the measurement goal.}  
For impact-oriented studies, ranking lists such as Tranco are appropriate because they reflect the security posture of highly visited domains. For prevalence-oriented studies, broad web archives such as Common Crawl provide access to a substantially larger and more diverse host population than conventional popularity lists. In our measurements, dataset choice changed observed prevalence by up to about 16 percentage points across reported header classes, e.g., Content-Type issues appear in 16.43\% of Common Crawl hosts but only 0.44\% of Tranco domains. Choosing the sampling strategy is therefore only part of the methodological decision; choosing the right source population is equally important.

\subsection{Limitations}
\label{sec:limitations}

Our study evaluates sampling strategies under realistic conditions and across security issues ranging from rare to common. While this diversity reduces the risk that our findings are tied to a particular class of vulnerabilities or defenses, we cannot claim full generalizability across all security properties, datasets, or domains. Also, our results rely on a large ground-truth baseline, with each domain crawled up to 250 pages. We believe this scale is sufficient for statistically meaningful comparisons, but substantially larger or structurally different populations could still reveal edge cases not captured in our dataset. Likewise, our measurements were performed from a fixed set of servers using standard browser automation. As in prior web measurement work, factors such as network locality, transient instability, CDN variability, bot detection, JavaScript timing, DNS resolution differences, and client-specific TLS behavior may affect what content and vulnerabilities are observed. Our fixed crawling depth also under-represents deep or authenticated content. These factors limit completeness and may add noise, but they are unlikely to change the comparative behavior of the sampling strategies we evaluate.

On top of that, our analysis of Common Crawl is limited to static security
issues because dynamic security analysis relies on live data, whereas
Common Crawl comprises static snapshots of web domains. However, if
researchers need to analyze dynamic issues, selecting domains from Common
Crawl and then analyzing the selected domains is an existing practice.
Prior work has combined these two steps. For example, Squarcina et
al.~\cite{squarcina2021can} used Common Crawl's web-graph PageRank scores
to select related domains and then performed live, browser-based security
analyses on those domains. Future work could explore how sampling may
affect dynamic security measurements by selecting domains from Common
Crawl, analyzing up-to-date content, and collecting data at a smaller scale. In addition, Common Crawl operates at a large scale and uses a search-engine style ranking based on links between web hosts; however, it still does not provide a complete snapshot of the entire Web. As such, our prevalence analysis may be limited by its coverage.

{Finally, our research mainly focuses on security issues, but our findings can also contribute to privacy impact.}
For the privacy issues, 
for example, HSTS misconfiguration enables protocol
downgrade attacks that expose user browsing history to network
eavesdroppers, and prior privacy work by Davitt et al.~\cite{davitt2024costrictor} studied the privacy implications of HSTS headers in Tor Browser. Similarly, TLS misconfigurations carry privacy risks beyond their security impact: Foppe et al.~\cite{foppe2018exploiting} have shown that certain TLS deployment flaws can leak identifiable information and enable user
tracking. These examples illustrate that 
sampling bias may carry implications for privacy-oriented
studies as well.  
Future work could extend the sampling analysis to privacy
metrics and examine whether the biases we observe in security
measurements generalizes to privacy problems.

\section{Related Work}
\label{sec:relatedwork}

\paragraph{Sample representativeness in web and security studies.} 
The representativeness of samples has been investigated by recent works. Zhang et al.~\cite{zhang2015sampling}, in their study on evaluating the accessibility of large websites, pointed out that sampling methods used in prior work may produce bias distributions of policy violations across websites. To address this issue, a new method called URLSamp is proposed, which clusters pages based on URL patterns for analysis, thereby effectively identifying accessibility issues present in webpages that originate from the same template. Tan et al.~\cite{tan2007WebpageUpdates} build an adaptive model leveraging historical update patterns and page popularity to infer which pages are most likely to change when given a sampled webpage and its change status. Their results show that it is most likely to find more updated webpages in the current or upper directories of the changed webpages. Beyond web related studies, Redmiles et al.~\cite{redmiles2019well} find that security and privacy researchers often rely on data collected from Amazon Mechanical Turk (MTurk) to evaluate security tools, understand users' privacy preferences and measure their online behavior. To check MTurk's performance, a comparison analysis was conducted between a probabilistic telephone sample and the MTurk sample with a census-representative web-panel. Decker et al.~\cite{decker2026vdpcollect} instead consider domains participating in vulnerability disclosure programs as the measurement population, and find that such sites exhibit better security practices and fewer vulnerabilities than popular domains. While the focus of that work is not on sampling strategies, both works demonstrate the impact of the dataset choice on the measurement.

\paragraph{Web measurement experiment setups.} 
The impact of configurations in web measurement has also been investigated. Jueckstock et al.~\cite{jueckstock2021towards} compare how different measurement tools and setups affect the results of the study. Similarly, Demir et al.~\cite{demir2023similarity} investigate how different browser configuration profiles affect measurement results. Some works focus on the impact of crawlers in results. Ahmad et al.~\cite{ahmad2020apophanies} analyzed differences between crawlers and how crawler selection impacts understanding of the web ecosystem. Aleksei et al.~\cite{stafeev2024sok} empirically evaluate web crawling algorithms used in measurement studies. Many studies also analyze the domain ranking lists. Pochat et al.~\cite{LePochat2019} build the Tranco ranking list for research purposes, which is designed to be resistant against manipulation. Xie et al.~\cite{xie2022building} investigate the issues in popular domain ranking and build a new voting-based domain ranking list.
 
\paragraph{Reproducibility of measurement studies.} Approaches to improve reproducibility have been proposed. Paxson et al.~\cite{paxson2004strategies} propose strategies for sound internet measurement based on their practical experience. Demir et al.~\cite{demir2022reproducibility} investigate the reproducibility and replicability of web measurement studies. These studies reveal that many studies fail to yield reproducible results, and even slightly different setups could result in different conclusions.

\section{Conclusion}
\label{sec:conclusion}
In this paper, we conducted the systematic analysis of sampling strategies in web security measurement research to analyze how choices of sampling strategy affect the study. We first performed a comprehensive literature review resulting in 107 papers adopting a variety of eight different sampling strategies. Then, we performed large-scale analysis to generate the ground truth of security issue distribution among two large-scale datasets.
Our systematic comparison reveals that Top N sampling, despite its popularity among related works and rationality in certain research with specific target datasets, introduces significant and persistent bias when used to estimate properties beyond its declared prefix, whereas probability-based strategies remain accurate and stable. Hybrid sampling, meanwhile, yields no advantage over probability sampling. We further reveal that the choice of dataset must be aligned with the research goal: a misaligned dataset may introduce bias of up to 19.6\%. Through a dedicated experiment with case studies, we provide suggestions for performing representative web measurement studies for researchers' reference, advocating adaptive probability-based sampling.


\bibliographystyle{ACM-Reference-Format}
\bibliography{sample-base}

\appendix
\section{Ethical Considerations}
\label[appendix]{sec:ethics}

This section provides more details about our ethical considerations of our research paper. 

\paragraph{Avoiding Overload of Web and Nameserver Infrastructure.}
Large-scale measurements risk generating excessive traffic to hosting providers and authoritative nameservers, particularly when many domains share the same backend infrastructure. To mitigate this, we distributed the 500{,}000 domains into 50 shuffled buckets, reducing sequential access patterns that could concentrate load on individual servers. Each machine processed at most two buckets concurrently, and our crawler limited each domain to a maximum of 250 pages, with a maximum crawl depth of two and a 30\,s timeout. These restrictions ensured that our traffic remained comparable to typical benign crawlers and avoided creating undue strain on shared infrastructures.

\paragraph{Avoiding Overload of Public DNS Resolvers.}
Because DNS-based preprocessing could generate significant load on the public resolvers of Google and Cloudflare, we applied strict controls on query rate. DNS lookups were issued in batches of 100 domains, with at most 10 concurrent queries and a mandatory 2\,s pause between batches to ensure a low sustained query rate. We observed no DNS lookup failures attributable to resolver rate limits or throttling, suggesting that our approach did not stress these services.

\paragraph{Non-intrusive Security Testing.}
Security measurements may unintentionally trigger server-side vulnerabilities or interfere with remote systems if they involve active exploitation attempts. Our methodology strictly avoided any server-side payload injection, fuzzing, or exploitation. All vulnerability detection—such as identifying client-side XSS or header misconfigurations—was performed solely through passive analysis of fetched page content, JavaScript execution in our controlled environment, and TLS inspection. Thus, our tests did not alter or probe server-side state beyond standard web browsing behavior.

\paragraph{Handling Sensitive Vulnerability Information.}
Our study identified approximately 100k domains exhibiting at least one security issue, which poses a risk of harm if sensitive data were released publicly. Due to the scale of the dataset, individually notifying all affected site operators is impractical. To prevent misuse, we publicly share only aggregated results that cannot be traced back to specific domains. Access to the full 500{,}000-domain dataset is granted solely through a formal request process restricted to researchers who provide a clear explanation of research goals, intended data use, and risk mitigation procedures. We have started the vulnerability-disclosure process. We first revisit affected websites to verify that each finding is still reproducible. For confirmed findings, we identify an official security or organizational contact and submit a concise report containing the relevant configuration evidence and remediation guidance.

\paragraph{Risks of Publishing Methodological Insights.}
Beyond concerns related to data collection, one might question whether disseminating our findings on effective sampling strategies introduces risks by enabling misuse of improved measurement methodologies. Our intended audience consists of researchers seeking to design more representative and reliable large-scale security studies. The knowledge we present focuses on methodological rigor, e.g., which sampling strategies yield stable measurements, not on how to exploit systems or identify specific vulnerable services. As such, our contributions do not provide actionable information for harming users, website operators, or infrastructure providers. Instead, they support reproducibility and robustness within the research community. We therefore assess the residual risk of publishing these methodological results as negligible and outweighed by the scientific value of improved measurement practices.

\section{Open Science}
\label[appendix]{sec:artifacts}


We provide a GitHub repository\footnote{\href{https://github.com/viewv/lsweb\_ccs26}{\texttt{https://github.com/viewv/lsweb\_ccs26}}}. This repository includes our tools for crawling and analyzing the security issues of domains in the Tranco ranking list, collecting the Common Crawl dataset, and identifying security header issues. We also release the reviewed paper list, the data and code for our case study, an example tool to help researchers apply our adaptive sampling strategy, and non-identifying results derived from our measurements. However, due to the concerns presented in \Cref{sec:ethics}, we provide restricted access to the complete domain dataset or the sampled domain sets for each strategy. The datasets are available only upon formal request from researchers who must describe their intended use, justify the necessity of accessing sensitive information, and outline appropriate safeguards.

\section{Identification of Header Misconfiguration}
\label[appendix]{sec:header}

\paragraph{Content injection protection:} Content Security Policy (CSP) has been crucial in preventing attackers from inserting unexpected scripts to trusted websites. In terms of misconfiguration, we specifically check if the CSP configuration in \textit{script-src} is missing or too permissive, i.e., allowing wildcard $*$, containing \texttt{unsafe-inline}, or \texttt{unsafe-eval}. We identify these attributes as security-sensitive misconfiguration because wildcard allows loading scripts from any sources, and \texttt{unsafe-inline} allows inline scripts which still can execute malicious code if developers not configured \texttt{nonce} or \texttt{hash} for the inline code. Similarly, the attribute \texttt{unsafe-eval} allows the use of \texttt{eval()} and similar methods, which can execute arbitrary code. On top of \textit{script-src} field, stylesheet scripts may also incur code injection attacks from the attackers~\cite{stylesheetinject}. Hence, we also consider whether \textit{style-src} field contains \texttt{unsafe-inline} or wildcard.

\paragraph{Clickjacking protection:} For clickjacking protection, we check whether clickjacking protection is not configured (i.e., neither \textit{X-Frame-Options} nor \textit{CSP frame-ancestors} is configured) or configured with invalid values (e.g., \textit{X-Frame-Options} not set to \texttt{DENY} or \texttt{SAMEORIGIN}).

\paragraph{Cookie security:} As a fundamental building block of CSRF prevention, we check the configuration on cookies by analyzing whether the \textit{SameSite} parameter or whether secure options are properly configured. For \textit{SameSite} attribute, we check if it is set to \texttt{None} without \texttt{Secure} attribute, which is an invalid configuration that may lead to cookie being exploited. For secure options, we check whether any of the three attributes \texttt{SameSite}, \texttt{Secure}, and \texttt{HttpOnly} are missing, as they allow Cross-site requests, transmission over insecure HTTP, and cookie access via JavaScript that can be fetched via XSS attack, separately.
    
\paragraph{HTTP Strict Transport Security:} For the configuration on HTTP Strict Transport Security (HSTS), we check whether the configuration is missing or max age is configured too short, because missing HSTS allows protocol downgrade and results in cookie hijacking over HTTP, and short max age may not provide sufficient protection duration. While prior work suggests HSTS max-age values from 180 days up to one year~\cite{hsts_chosen,hsts_maxage,hsts_owasp,testssl}, we set our threshold at 180 days to focus on configurations that are unambiguously insecure, avoiding the gray area between 180 days and one year to maintain a conservative and reliable baseline.

\paragraph{Cross-Origin Resource Sharing:} Cross-Origin Resource Sharing (CORS) misconfigurations can lead to sensitive data exposure. We specifically check for the misconfiguration by identifying whether two criteria are met. First, \textit{Access-Control-Allow-Origin} is set to wildcard, allowing resources to be accessed from any origin. Second, \textit{Allow-Credentials} is set to true, allowing cross-origin to include credentials such as Cookies, which may lead to sensitive data exposure.

\paragraph{Content Type and MIME Handling:} As an essential part of the HTTP protocol, Content Type and MIME Handling configuration indicates the media type of resources being sent. In terms of misconfiguration, we care about whether the \textit{content-type} is missing which leads to potential misinterpretation, or whether \textit{X-Content-Type-Options} is set to an invalid value (i.e., not set to \texttt{nosniff}) which allows potential MIME sniffing attacks.

\section{Breakdown of Error Types at Collection Time}
\label[appendix]{app:errortypes}
\begin{table}[t]
\centering
\footnotesize
\setlength{\tabcolsep}{3pt}

\begin{tabular}{l l r}
\toprule
\textbf{Category}&\textbf{Error Type} & \textbf{Count} \\
\midrule

Network& Unknown Host & 64,447 \\
Timeout& Exceeded 30000ms & 32,135 \\
Network& Connection Refused& 6,040 \\
Network& Network Reset & 2,993 \\
Network& Binding Aborted & 1,316 \\
Timeout& Network Timeout & 404 \\
Unknown& SSL Unknown Error & 340 \\
Network& Redirect Loop & 299 \\
Network& Error Abort & 271 \\
Unknown& General Unknown Error & 261 \\

\bottomrule
\end{tabular}
\caption{Error Types and Their Frequencies}
\label{tab:error_types}
\end{table}

During the experiment on Tranco 500k, we identified roughly 20\% of the domains in the list that failed to return result during crawling. To quantify their impact, we analyzed all encountered errors and summarized the top 10 most frequent types in \Cref{tab:error_types}. Among them, network issues dominate these issues, accounting for 6 of the top 10 errors. The most common one is failure to resolve the A record of a domain name, which may occur when the website owners configure their websites incorrectly, resulting in failure to resolve the domain that prevents us from crawling the data from the website. Timeout and connection-refused errors are the next two most frequent ones, reflecting servers that either did not respond within 30 seconds or actively rejected connections.

\begin{figure}
    \centering
    \hspace*{-5mm}
    \includegraphics[width=\linewidth]{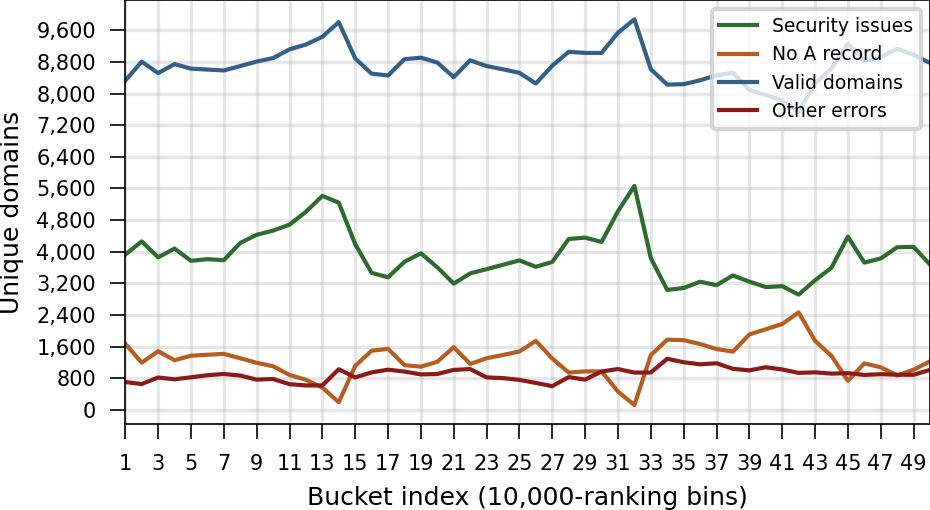}
    \caption{Distribution of errors, accessibility, and vulnerability of crawled domains in 10k ranking bins.}
    \label{fig:security_vs_errors}
\end{figure}

\Cref{fig:security_vs_errors} shows counts of valid domains and detected issues across 50 buckets of 10{,}000 domains ordered by Tranco rank. While error rates (solid red and orange) vary across buckets, the share of valid domains (solid blue) consistently remains between 72\% and 96\%. Moreover, domains with security issues (solid green) exhibit a non-uniform distribution, confirming that our dataset still meaningfully reflects real-world variation and remains appropriate for evaluating sampling-strategy behavior.

\section{Shapley Value Formulation for Hybrid Sampling}
\label[appendix]{sec:shapley_detailed}

This appendix provides additional detail on how we formulate the component analysis of hybrid sampling using Shapley values. The main paper presents only the conceptual intuition and high-level interpretation; here, we include the formal definitions used to compute the marginal contribution of each component in a hybrid sampler.

\subsubsection*{General Shapley Value Definition}

Let \(N\) be the set of players and \(v(S)\) the value (or payoff) associated with any coalition 
\(S \subseteq N\). The Shapley value of player \(i \in N\) is:

\[
\phi_i
= \sum_{S \subseteq N \setminus \{i\}}
\frac{|S|! \, (|N|-|S|-1)!}{|N|!}
\left[ v(S \cup \{i\}) - v(S) \right].
\]

The combinatorial weight
\[
\frac{|S|!\,(|N|-|S|-1)!}{|N|!}
\]
ensures that each ordering of players joining a coalition receives equal probability.

\subsubsection*{Two-Component Hybrid Sampling as a Shapley Game}

In our setting, a hybrid sampling algorithm consists of two components:

\begin{itemize}
    \item \textbf{A}: the Top N deterministic prefix (e.g., Top 5k), and
    \item \textbf{B}: the rest-sampling mechanism (Random, Bucket, or Stratified).
\end{itemize}

Thus the player set is simply: $ N = \{A, B\}$. With two players, there are only two possible joining orders, and the Shapley formula reduces to:

\[
\phi_A = \frac{1}{2}\big[ v(A) - v(\varnothing) \big]
    + \frac{1}{2}\big[ v(A,B) - v(B) \big],
\]

\[
\phi_B = \frac{1}{2}\big[ v(B) - v(\varnothing) \big]
    + \frac{1}{2}\big[ v(A,B) - v(A) \big].
\]

As is common in error-based applications, we set \(v(\varnothing)=0\).

\subsubsection*{Value Function}

To quantify estimation accuracy, we define the value of any coalition as the negated expected error:

\[
v(S) = -\,\mathbb{E}\!\left[ \big| \text{error}(S) \big| \right].
\]

Higher values correspond to lower estimation error.  
Using this definition, the relevant coalition values become:
\[
v(A) = -E_A,\qquad
v(B) = -E_B,\qquad
v(A,B) = -E_{AB},
\]

where \(E_A\), \(E_B\), and \(E_{AB}\) denote the Top N-only, Rest-only, and Hybrid errors, respectively Substituting these into the two-player Shapley equations yields:
\[
\phi_A = \tfrac{1}{2}(-E_A)
        + \tfrac{1}{2}(-E_{AB} + E_B),
\qquad
\phi_B = \tfrac{1}{2}(-E_B)
        + \tfrac{1}{2}(-E_{AB} + E_A).
\]

A positive Shapley value indicates that the corresponding component reduces estimation error; a negative value indicates that it increases error.

\subsubsection*{Experimental Procedure}

For each security issue, we compute Shapley values using the following workflow:

\begin{enumerate}
    \item Compute the ground-truth prevalence over the full dataset.
    \item Fix a Top N cutoff to define component \(A\).
    \item For a range of rest-sample sizes \(n_{\text{rest}}\):
    \begin{enumerate}
        \item draw rest-samples using Random, Bucket, or Stratified sampling,
        \item measure rest-only error \(E_B\),
        \item measure the hybrid error \(E_{AB}\) by combining Top N with the rest-sample.
    \end{enumerate}
    \item Compute \(\phi_A\) and \(\phi_B\) using the equations above.
\end{enumerate}

\subsubsection*{Interpretation} 

A positive value indicates that adding the component reduces estimation error on average, whereas a negative value means that the component increases error. Thus, \(\phi_A > 0\) means the Top N block contributes beneficially, while \(\phi_A < 0\) indicates that it introduces bias or distortion. Likewise, \(\phi_B > 0\) shows that the rest-sampling mechanism improves estimation, and \(\phi_B < 0\) signals that it worsens accuracy.

\section{Sampling Strategies Details}
\label[appendix]{sec:appendix_sampling_strategies}

This appendix provides additional detail on how we implement the sampling strategies discussed in \Cref{sec:sample}. The Top N strategy and Random N strategy are straightforward, so we will focus instead on the Systematic sampling and Stratified sampling strategies.

\subsubsection*{Systematic sampling}

Systematic sampling involves selecting samples at fixed intervals $k$, starting from a random starting point $s$ in an ordered list of $N$ data points. As we could not find prior work in web measurement offering a reference for the calculation and use of fixed intervals, we used the fractional sampling interval approach~\cite{iachan1982systematic} discussed by Kish et al.~\cite{kish1965survey} and Murthy et al.~\cite{murthy1967sampling}. 

The fractional sampling interval first selects the element at a random position \( s \in [1, k) \). Then, it uses a real fixed interval \( k = N/n \), where $N$ is the total sample set and $n$ is the desired sample size. The index of the $i$-th element is then computed as:

\[
\text{idx}_i = \left\lfloor s - 1 + i \cdot \frac{N}{n} \right\rfloor \bmod N + 1
\]

The floor operation ensures the index is an integer, whereas the modulo operation prevents index overflow. In our implementation, we generate a random starting position $s$ using a uniform distribution over the first interval.

\subsubsection*{Stratified random sampling}

Stratified random sampling first partitions the population into strata (subgroups) according to relevant characteristics. A sample is then drawn from each stratum in proportion to the size of the stratum in the overall population, and the selected elements are combined to form the final sample. In web measurement studies, strata are often constructed based on website rankings, where lower-ranked websites typically form larger strata. In this work, we adopt the stratification configuration proposed by Demir et al.~\cite{demir2023similarity}. The strata are defined in~\Cref{tab:stratified-layers}.

\begin{table}[t]
\centering
\footnotesize
\begin{tabular}{c l}
\toprule
\textbf{Strata} & \textbf{Rank Range} \\
\midrule
1   & [1, 5{,}000]          \\
2   & [5{,}001, 10{,}000]    \\
3   & [10{,}001, 50{,}000]   \\
4   & [50{,}001, 250{,}000]  \\
5   & [250{,}001, 500{,}000] \\
\bottomrule
\end{tabular}
\caption{Stratified stratum configurations based on website rankings.}
\label{tab:stratified-layers}
\end{table}

For a given total sample size~$n$, we allocate samples to strata using proportional allocation. Let the strata be indexed by $i \in \{1,\dots,5\}$, and let $N_i$ denote the population size of stratum~$i$ (i.e., the number of domains within the corresponding rank interval). The total population size is $N=\sum_{i=1}^{5} N_i$. The ideal (fractional) allocation for stratum~$i$ is

\[
t_i = n \cdot \frac{N_i}{N}.
\]

We then convert the fractional allocations into integer sample counts. First, we take the floor of each value:

\[
a_i = \lfloor t_i \rfloor,
\qquad
A = \sum_{i=1}^{5} a_i.
\]

If $A < n$, a remainder of 
\[
r = n - A
\]

samples must still be assigned. Following adjustment practice, we distribute these $r$ samples to the strata with the largest $N_i$, since the largest strata should have more elements. After obtaining the integer allocation vector $\mathbf{a} = (a_1,\dots,a_5)$, we randomly draw $a_i$ samples uniformly from each stratum~$i$. Finally union all selected elements constitutes one stratified random sample of size~$n$. 

In addition, the buckets sampling strategy can be seen as a special case of stratified random sampling, where each stratum contains an equal number of elements, we adopt the same allocation and sampling procedure as described above and we use the buckets numbers 10 proposed by Hantke et al.~\cite{hantke2023you}.

\section{Adaptive Sampling strategy}
\label[appendix]{sec:adaptive-sampling}

Researchers generally do not know the affected-unit rate before conducting a
measurement. A fixed sample may therefore be unnecessarily large for a common
issue but insufficient for a sparse one. To make the recommendation in
\Cref{sec:bestpractice} actionable, this appendix specifies the adaptive
probability sampling strategy and evaluates it in a case study.

\subsection{The Adaptive Strategy}

Let $\mathcal{F}$ be the declared population containing $N$ units. We draw one
random permutation of $\mathcal{F}$ and inspect nested prefixes of that
permutation. Consequently, sampling is uniform and without replacement, and
each new stage retains all units measured at previous stages while drawing
only the additional units from the remaining population. For example, we
expand the sample by doubling, giving cumulative sampling fractions of 2\%,
4\%, 8\%, 16\%, 32\%, 64\%, and, if necessary, 100\%. The 2\% pilot is not 
allowed to stop because no preceding estimate is available for comparison.

At each stage, the researcher performs the following steps.

\paragraph{Step 1: Estimate the affected-unit rate.}
Let $n_k$ be the cumulative number of inspected units at stage $k$, and let
$x_k$ be the number of affected units observed among them. The current estimate
is

\begin{equation}
    \hat{p}_k = \frac{x_k}{n_k}.
\end{equation}

If $x_k=0$, the procedure does not stop because the available sample contains
insufficient information about the frequency of the issue. Instead, the
cumulative sample size is doubled and the next stage is evaluated.

\paragraph{Step 2: Quantify the uncertainty of the current estimate.}
If at least one affected unit has been observed, we calculate the half-width
of the approximate confidence interval for a proportion. For example, with a
95\% confidence level, the half-width is given by

\begin{equation}
    h_k =
    1.96
    \sqrt{
        \frac{\hat{p}_k(1-\hat{p}_k)}{n_k}
        \frac{N-n_k}{N-1}
    }.
    \label{eq:adaptive-ci}
\end{equation}

The value 1.96 corresponds to the 95\% confidence level. The factor
$(N-n_k)/(N-1)$ is the finite-population correction: it accounts for sampling
without replacement, so the uncertainty decreases as more of the population
is inspected and becomes zero when the complete population is measured.

\paragraph{Step 3: Check stability and precision.}
Starting from the second stage, we compare the current estimate with the estimate
from the preceding stage. We calculate

\begin{equation}
    D_k =
    \frac{|\hat{p}_k-\hat{p}_{k-1}|}{\hat{p}_k},
    \qquad
    R_k =
    \frac{h_k}{\hat{p}_k}.
    \label{eq:adaptive-stop}
\end{equation}

Here, $D_k$ is the relative change between two consecutive estimates. It
checks whether doubling the sample materially changes the result. In contrast,
$R_k$ is the relative half-width of the current 95\% confidence interval. It
checks whether the current estimate has sufficient statistical precision.
These conditions serve different purposes: two consecutive estimates may be
similar by chance even when both are imprecise, so the relative-change
condition alone is insufficient, while the confidence interval describes
uncertainty at the current stage but does not verify that the estimate
remained stable after expanding the sample. We therefore require both
conditions.

\paragraph{Step 4: Stop or expand the sample.}
We use a user-specified relative tolerance $\tau$, for example we set to $0.05$ in
our evaluation. The strategy stops at stage $k$ only when

\begin{equation}
    x_k > 0,
    \qquad
    D_k \leq \tau,
    \qquad
    R_k \leq \tau.
    \label{eq:adaptive-decision}
\end{equation}

In other words, the strategy stops only when doubling the sample changes the
estimated rate by at most $\tau$ (5\% in our evaluation) and the half-width of
the current 95\% confidence interval is at most $\tau$ relative to the
estimate. If either condition is not satisfied, the cumulative sample size is
doubled and the next stage is evaluated.

\begin{algorithm}[t]
  \DontPrintSemicolon
  \caption{Adaptive probability sampling}
  \label{alg:adaptive-rate}
  \KwIn{Declared population $\mathcal{F}$ of size $N$; tolerance $\tau=0.05$}

  Draw one random permutation of $\mathcal{F}$\;
  $n_1 \leftarrow \lceil 0.02N \rceil$\;
  Inspect the first $n_1$ units and compute
  $\hat{p}_1 \leftarrow x_1/n_1$\;

  \For{$k \leftarrow 2,3,\ldots$}{
    $n_k \leftarrow \min(2n_{k-1},N)$\;
    Inspect only the additional units in positions
    $n_{k-1}+1,\ldots,n_k$\;
    $\hat{p}_k \leftarrow x_k/n_k$\;

    \If{$x_k>0$}{
      Compute the 95\% CI half-width $h_k$ using
      Equation~\ref{eq:adaptive-ci}\;
      $D_k \leftarrow
      |\hat{p}_k-\hat{p}_{k-1}|/\hat{p}_k$\;
      $R_k \leftarrow h_k/\hat{p}_k$\;

      \If{$D_k\leq\tau$ \textbf{and} $R_k\leq\tau$}{
        \textbf{return} $(\hat{p}_k,n_k)$\;
      }
    }

    \If{$n_k=N$}{
      \textbf{return} $(\hat{p}_k,n_k)$\;
    }
  }
\end{algorithm}

\paragraph{Choosing the tolerance.}
The tolerance $\tau$ is a user-specified precision parameter rather than a
fixed property of the strategy. We use $\tau=0.05$ in our main evaluation. A
study with a more limited measurement budget may select a looser tolerance,
such as $\tau=0.10$, whereas a study requiring greater precision may select a
smaller value. A smaller $\tau$ requires both greater agreement between
consecutive estimates and a narrower confidence interval, and therefore
generally requires a larger sample; a larger $\tau$ permits earlier stopping
but produces a less precise estimate. Researchers should predeclare the
tolerance according to the precision required by their research question and
report it together with the final sample size and confidence interval.

\subsection{Adaptive Sampling Strategy: Case Study}
\label[appendix]{sec:adaptive-case-study}

We evaluate whether the adaptive strategy stops at an appropriate stage without
access to the population-wide rate. We use three outcomes representing high,
medium, and low affected-unit rates: Cookie security, Clickjacking, and CORS.
We apply the same strategy to three declared populations: the Tranco Top 100K,
the Tranco Top 500K, and the Common Crawl hosts. The
Top 100K experiment models a resource-constrained study whose question concerns
exactly the Top 100K domains. The Common Crawl experiment models a
prevalence-oriented study over a much broader host population.
Table~\ref{tab:adaptive-case-studies} in \Cref{sec:bestpractice} summarizes
the results; the following paragraphs provide the detailed per-population
analysis.

A researcher would run one adaptive sequence in a real measurement. To
evaluate the reliability of that one-run strategy, we repeat
the experiment over 1{,}000 independently seeded random permutations of each
known population. The adaptive stopping function receives only the
observations available at its current stage. In particular, it does not
receive the full-population rate. We use the latter only after stopping to
calculate the relative error

\begin{equation}
    E =
    \frac{|\hat{p}_{\mathrm{stop}}-p_{\mathcal{F}}|}
         {p_{\mathcal{F}}},
\end{equation}

where $p_{\mathcal{F}}$ is the affected-unit rate obtained from the complete
population. We define retrospective reliability as the fraction of the 1{,}000
trials for which $E\leq0.05$.

\paragraph{Top 100K results.}
For Cookie security, 98.3\% of the trials stop at 8\%, and the remaining 1.7\%
stop at 16\%. The mean sample is 8.14\% of the population, and 99.0\% of the
stopped estimates are within 5\% relative error of the complete-population
rate. Clickjacking stops at 16\% in 97.7\% of trials and at 32\% in the
remaining 2.3\%, producing 98.3\% retrospective reliability with a mean sample
of 16.37\%.

The low-frequency CORS outcome behaves differently. Its full-population rate
is only 0.759\%, and the strategy reaches the complete Top 100K in every
trial. Although this setting provides no measurement-cost reduction, it
prevents the study from reporting an estimate that does not satisfy the
requested precision. For comparison, only 10.4\% of fixed 2\% samples and
21.9\% of fixed 4\% samples are within the same 5\% relative-error tolerance,
and a rule based only on cross-stage change reaches 46.5\%. The adaptive
result therefore communicates an actionable limitation: accurately estimating
this sparse outcome within the small Top 100K population requires measuring
substantially more than the initially planned budget.

\paragraph{Top 500K results.}
For the larger Tranco population, Cookie security stops at 4\% in every trial,
and Clickjacking stops at 4\% in 97.6\% of trials. Their mean sampling
fractions are 4.00\% and 4.10\%, with respective reliabilities of 99.9\% and
99.2\%. Sparse CORS requires considerably more data: 97.2\% of trials stop at
64\%, and the mean sampling fraction is 65.01\%. This confirms that a common
issue can be estimated economically, whereas a strict relative-precision
target for a rare issue can require most of the declared population.

\paragraph{Common Crawl results.}
On the Common Crawl dataset, Cookie security and
Clickjacking stop at 4\% in every trial. CORS stops at 4\% in 99.4\% of trials
and at 8\% in 0.6\%. The corresponding reliabilities are 100.0\%, 100.0\%, and
99.0\%. Although the CORS rate is lower in Common Crawl than in either Tranco
population, a 4\% Common Crawl sample contains many more positive observations
in absolute terms because the population is much larger. The stopping
behavior therefore depends jointly on the affected-unit rate, population size,
and requested precision, rather than on the percentage alone.

\end{document}